\documentclass[preprint,12pt]{elsarticle}
\usepackage[margin=1in]{geometry}
\usepackage{setspace}
\usepackage[T1]{fontenc}
\usepackage{lmodern}
\usepackage{amsmath,amssymb,mathtools,bm,graphicx,booktabs,array,tabularx,multirow}
\usepackage[lofdepth,lotdepth]{subfig}
\usepackage{float}
\usepackage{url}
\usepackage{lineno,longtable,placeins,microtype,multicol}
\usepackage{algorithm,algpseudocode}
\usepackage[hidelinks]{hyperref}
\hypersetup{
  pdftitle={An Invariant Neural-Network Eddy-Viscosity Closure with Iterative Solver Coupling for Molten Salt Fast Reactor Thermal-Hydraulics},
  pdfauthor={Zaid Abulawi; William Oliveira; Akzhol Almukhametov; Jun Fang; Ling Zou; Rui Hu; Stefano Lorenzi; Yang Liu}
}
\makeatletter
\def\ps@pprintTitle{%
  \let\@oddhead\@empty
  \let\@evenhead\@empty
  \def\@oddfoot{\@myfooterfont\hfil\@elsarticlemyfooter\hfil}%
  \let\@evenfoot\@oddfoot
}
\makeatother
\newcolumntype{Y}{>{\raggedright\arraybackslash}X}
\algrenewcommand\algorithmicrequire{\textbf{Input:}}
\algrenewcommand\algorithmicensure{\textbf{Output:}}
\journal{Physics of Fluids}
\myfooter[C]{}

\begin{document}

\begin{frontmatter}

\title{An Invariant Neural-Network Eddy-Viscosity Closure with Iterative Solver Coupling for Molten Salt Fast Reactor Thermal-Hydraulics}

\author[nuen]{Zaid Abulawi}

\author[polimi]{William Oliveira}

\author[nuen]{Akzhol Almukhametov}

\author[anl]{Jun Fang}

\author[anl]{Ling Zou}

\author[anl]{Rui Hu}

\author[polimi]{Stefano Lorenzi}

\author[nuen]{Yang Liu\corref{cor1}}
\ead{y-liu@tamu.edu}

\address[nuen]{Department of Nuclear Engineering, Texas A\&M University, College Station, TX 77843, USA}
\address[polimi]{Department of Energy, Nuclear Engineering Division, Politecnico di Milano, Milan, Italy}
\address[anl]{Nuclear Science and Engineering Division, Argonne National Laboratory, Lemont, IL 60439, USA}

\cortext[cor1]{Corresponding author.}

\begin{abstract}
System thermal-hydraulics codes such as the System Analysis Module (SAM) close turbulence on coarse meshes with algebraic zero-equation mixing-length models. These models cannot represent the non-equilibrium turbulence, flow separation, and large-scale recirculation of open-cavity advanced reactors. We develop an invariant neural-network eddy-viscosity closure for SAM and apply it to the Molten Salt Fast Reactor (MSFR). The closure predicts the turbulent kinetic energy and timescale of each element from the flow and geometry of its neighbors, without solving turbulence transport equations. The resulting non-negative eddy viscosity enters the implicit diffusive operator, which keeps the momentum equations well conditioned. We prove that the closure is exactly invariant to translation, Galilean boost, rotation, reflection, and neighbor ordering. A network trained on high-order Nek5000 fields sees a different feature distribution on the coarse SAM mesh. We remove this covariate shift by iteratively retraining the network on SAM flow states against fixed Nek5000 targets. For a held-out steady state, the coupled closure reproduces the single recirculation cell and eddy-viscosity plume of the Nek5000 reference. In a 120~s pump-coastdown transient, it halves the velocity error of the mixing-length model, from $37.6\%$ to $17.7\%$, and matches the Nek5000 eddy viscosity with a time-averaged relative error of $12\%$. The remaining velocity error comes from the coarse mesh. The coupled transient takes about 0.2 core-hours, compared with about 330 core-hours for Nek5000. The results show that the closure can recover two-equation RANS turbulence behavior in a coarse-mesh system code at a small fraction of the CFD cost.
\end{abstract}


\end{frontmatter}

\section*{Nomenclature}
\label{sec:nomenclature}
\begingroup
\small
\setstretch{1.0}
\raggedright
\setlength{\parskip}{1.2pt}
\setlength{\columnsep}{2em}
\newlength{\nomw}\setlength{\nomw}{3.6em}
\newcommand{\nom}[2]{\par\noindent\hangindent=\nomw\hangafter=1%
  \sbox0{#1}\ifdim\wd0>\dimexpr\nomw-0.7em\relax\usebox0\enspace\else\makebox[\nomw][l]{\usebox0}\fi#2}
\newcommand{\nomhead}[1]{\par\vspace{0.6ex}\noindent\textit{#1}\par\vspace{0.2ex}}
\begin{multicols}{2}
\nomhead{Physical quantities}
\nom{$A$}{element area [m$^2$]}
\nom{$d_w$}{wall distance [m]}
\nom{$k$}{turbulent kinetic energy [m$^2$\,s$^{-2}$]}
\nom{$L$, $T$}{characteristic length [m] and time [s]}
\nom{$N_e$}{number of mesh elements}
\nom{$p$}{effective mean pressure [Pa]}
\nom{$r$, $z$}{radial and axial coordinates [m]}
\nom{$Re$}{Reynolds number}
\nom{$\mathbf{S}$}{mean strain-rate tensor [s$^{-1}$]}
\nom{$t$}{time [s]}
\nom{$U$}{velocity scale or inlet speed [m\,s$^{-1}$]}
\nom{$\mathbf{u}$}{mean velocity vector [m\,s$^{-1}$]}
\nom{$V$}{axisymmetric element volume [m$^3$]}
\nom{$\mu$}{dynamic viscosity [Pa\,s]}
\nom{$\rho$}{fluid density [kg\,m$^{-3}$]}
\nom{$\tau$}{turbulence timescale [s]}
\nom{$\Omega$}{fluid domain}
\nomhead{Neural-network and closure quantities}
\nom{$\mathcal{A}$}{aggregated SAM feature dataset}
\nom{$b$}{wall-adjacency indicator}
\nom{$\mathbf{C}$, $\mathbf{c}$}{scalar feature matrix and row}
\nom{$\mathcal{D}$}{invariant descriptor matrix}
\nom{$E$}{relative $L_2$ error}
\nom{$\mathcal{F}_\Theta$}{neural closure mapping}
\nom{$\mathbf{G}$}{embedding matrix}
\nom{$m$}{number of embedding channels}
\nom{$n$, $\mathcal{N}$}{neighbor count and neighborhood}
\nom{$\mathbf{Q}$, $\mathbf{q}$}{feature matrix and row}
\nom{$\mathbf{r}$, $\mathbf{w}$}{relative position and velocity vectors}
\nom{$\mathbf{z}$}{transformed target vector, $(z_k, z_\tau)^{\mathsf{T}}$}
\nom{$\Theta$}{network parameters}
\nom{$\phi$}{neural network}
\nom{$\psi$}{normalization map}
\nomhead{Subscripts and superscripts}
\nom{$a$, $e$}{element indices}
\nom{$0$}{target element; characteristic scale}
\nom{$c$, $z$}{features and targets}
\nom{emb, pred}{embedding and prediction networks}
\nom{in, pk}{inlet and peak}
\nom{$k$, $\tau$}{target components}
\nom{$t$}{turbulent}
\nom{$w$}{wall}
\nom{${}^*$}{Nek5000 reference}
\nom{$\widehat{\cdot}$}{network prediction}
\nom{$\tilde{\cdot}$}{transformed frame}
\nom{${}'$}{reduced}
\nom{$\overline{\cdot}$}{domain or time mean}
\nom{${}^{\mathsf{T}}$}{transpose}
\end{multicols}
\endgroup

\section{Introduction}
\label{sec:intro}
Liquid-fueled molten salt reactors (MSRs) dissolve fissile and fertile actinides in a circulating fluoride or chloride carrier salt~\cite{ROUCH2014449,rubiolo2017high,brovchenko2019neutronic}. The Molten Salt Fast Reactor (MSFR) is the MSR concept selected by the Generation-IV International Forum. The fuel salt generates fission heat and carries delayed-neutron precursors through an open core cavity without coolant channels. Reactor neutronics and safety margins therefore depend strongly on the multidimensional core flow~\cite{AUFIERO2014390,di2022multiphysics,mui2025integrated}. The MSFR core has curved inner shield walls, an abrupt expansion from the cold-leg inlet, and strong centripetal acceleration toward the hot-leg outlet. At Reynolds numbers above $10^6$, this geometry produces shear-layer separation, impinging jets, and persistent recirculation zones. These zones govern residence times, wall heat transfer, and peak fuel temperatures in normal operation and during pump-coastdown transients~\cite{ROUCH2014449,LIU2022112059}. Through thermal and precursor-transport feedback, they also shape the core reactivity~\cite{AUFIERO2014390,di2022multiphysics}. A flow model that misrepresents them can mispredict the power and temperature evolution, including hot spots in low-velocity regions near the walls~\cite{ROUCH2014449}.

Simulating these multidimensional phenomena over long plant transients requires a balance between physical fidelity and computational cost~\cite{long2021review,petruzzi2008thermal}. High-fidelity computational fluid dynamics (CFD) codes such as Nek5000~\cite{Merzari01092020,fang2021fullcore}, OpenFOAM~\cite{AUFIERO2014390,di2022multiphysics}, and commercial finite-volume packages~\cite{fang2024cfd} resolve the MSFR cavity on fine boundary-layer meshes. They use two-equation Reynolds-averaged Navier--Stokes (RANS) models, large-eddy simulation (LES), or direct numerical simulation (DNS)~\cite{zhiyin2015large,alfonsi2011direct,Tomboulides2025KTau}. However, their fine meshes and small time steps make them too expensive for routine plant-level transient sweeps, sensitivity studies, and uncertainty quantification~\cite{liu2019uncertainty,liu2021uncertainty,liu2023uncertainty,abulawi2025bayesian,wu2025uncertainty,abulawi2026bayesian}. Conversely, system and coarse-mesh multidimensional codes such as the System Analysis Module (SAM)~\cite{hu2021sam,hu2025sam,zou2020development} are fast because they use coarse finite-element meshes and algebraic zero-equation mixing-length closures~\cite{bradshaw1974possible,hu2025sam_manual}. System codes in general rely on empirical closures calibrated against separate-effect tests in simple tube and channel geometries~\cite{liu2021experimental}. Mixing-length closures work well in simple pipe and channel flows. They assume local equilibrium between turbulent production and dissipation and set the mixing length from the wall distance alone. In the open MSFR cavity, this local assumption misses non-local history and transport effects and produces qualitatively incorrect core recirculation patterns.

Data-driven turbulence modeling can transfer high-fidelity closure physics into coarse-mesh solvers without adding stiff transport partial differential equations (PDEs) at runtime~\cite{duraisamy2019turbulence,beck2019deep,KURZ2023109094,zhu2019machine,taghizadeh2021turbulence,Zhang_review,Girimaji2024TurbulenceClosureML,lanade2025high,Oulghelou2025MLBlending,toma2026mixed,Deanesi2023}. Early machine-learning RANS models trained neural networks or random forests to predict Reynolds-stress anisotropy or optimal eddy viscosity from Pope's integrity basis of strain-rate and rotation-rate invariants and other mean-flow features~\cite{Pope1975,ling2016reynolds,wu2017priori,yin2020feature,mcconkey2022deep}. These models can be accurate \textit{a priori} on held-out CFD snapshots. Inside a momentum solver, \textit{a posteriori}, they face three challenges~\cite{Wu2019IllConditioned,Liu2021IterativeML,bhushan2023assessment}.

First, the coupled momentum equations must remain well conditioned. Wu et al.~\cite{Wu2019IllConditioned}, Liu et al.~\cite{Liu2021IterativeML}, and McConkey et al.~\cite{mcconkey2022deep} showed that a learned Reynolds stress injected explicitly as a source term can make the discretized RANS system severely ill-conditioned at high Reynolds numbers. Small regression errors then grow into large mean-velocity errors unless a non-negative eddy viscosity is treated implicitly.

Second, the closure needs non-local inputs that respect the symmetries of the flow. Furthermore, purely local strain-rate mappings $\mu_t = f(\mathbf{S})$ become multi-valued in separated or recirculating flows, because regions with similar local strain rates carry different turbulence levels~\cite{Liu2021IterativeML,man2024nonunique}. A constitutive closure must also be Galilean invariant under uniform boosts and frame invariant under translations, rotations, and reflections~\cite{Speziale1998,ling2016reynolds,yin2020feature,kaszuba2025implicit,berrone2022invariances}. Non-local closures built on sampled point sets represent spatial transport on fine CFD grids~\cite{ZHOU2021113927,ZHOU2022114211,Xu2022PDEFree}. Coarse-mesh finite-element solvers such as SAM and the underlying MOOSE framework~\cite{lindsay2021automatic,zou2020development,hu2025sam} have too few, variably sized elements for point-set sampling. A non-local closure for these solvers must instead gather flow and geometry data from neighboring elements. It must also be independent of the order of those neighbors and insensitive to how many there are.

Third, discretization differences cause covariate shift. A network trained on velocity and strain-rate fields projected from a fine, high-order spectral-element solver (Nek5000) sees a different feature distribution when it runs inside a coarse, first-order stabilized finite-element solver (SAM). Recent iterative and field-inversion RANS studies show that this shift degrades \textit{a posteriori} stability and accuracy unless the training adapts to the flow states the solver actually produces~\cite{Liu2021IterativeML,taghizadeh2021turbulence,fang2024field}.

Closest to the present work, Cervi et al.~\cite{cervi2026sam} added neural-network eddy-viscosity closures trained on CFD data to SAM to model thermal mixing and stratification in liquid-metal fast reactor plena. They found that transformer-based models outperform convolutional networks. The present closure differs in two ways. It is invariant to frame changes and neighbor ordering by construction, and it is retrained iteratively on SAM flow states to remove the covariate shift.

This work develops an invariant neural-network eddy-viscosity closure for SAM and applies it to steady-state and transient MSFR flows. Building on high-to-low data-driven modeling for advanced reactor flows~\cite{LIU2022112059,lanade2025high}, the closure takes its inputs from a two-layer neighborhood of mesh elements in SAM and MOOSE~\cite{zou2020development,hu2025sam}. This PDE-free closure predicts the turbulent kinetic energy and turbulence timescale of each element and forms a non-negative eddy viscosity from them, without solving transport PDEs. The network is trained iteratively with SAM~\cite{Liu2021IterativeML,fang2024field}. Section~\ref{sec:invariance_proofs} shows that the closure satisfies the required frame and permutation invariances exactly. Section~\ref{sec:iterative_theory} shows how the closure and its iterative training address each of the three challenges. We evaluate the closure \textit{a priori} on held-out Nek5000 snapshots and \textit{a posteriori} inside SAM, for steady states over a range of Reynolds numbers and for a 120~s pump-coastdown transient. SAM's mixing-length model serves as the baseline.

The remainder of this paper is organized as follows. Section~\ref{sec:background} presents the governing equations and baseline turbulence closures in Nek5000 and SAM. Section~\ref{sec:methodology} details the neural closure and the iterative training. Section~\ref{sec:test_case} describes the detailed MSFR case study. Section~\ref{sec:results} presents the \textit{a priori} and \textit{a posteriori} evaluation results and iterative convergence analysis, and Section~\ref{sec:conclusions} summarizes the main conclusions.

\section{Governing Equations and Baseline Turbulence Closures}
\label{sec:background}

Both the high-fidelity reference solver (Nek5000) and the coarse-mesh target solver (SAM) model turbulent incompressible and low-Mach-number single-phase flows via the Reynolds-averaged conservation of mass and momentum:
\begin{align}
    \nabla \cdot (\rho \mathbf{u}) &= 0, \label{eq:continuity} \\
    \rho \frac{\partial \mathbf{u}}{\partial t} + \rho (\mathbf{u} \cdot \nabla)\mathbf{u} &= -\nabla p + \nabla \cdot \!\left[ (\mu + \mu_t)\!\left(\nabla \mathbf{u} + \nabla \mathbf{u}^{\mathsf{T}}\right) \right] + \rho \mathbf{g}, \label{eq:momentum}
\end{align}
where $\mathbf{u}$ is the mean velocity vector in the two-dimensional axisymmetric meridional domain $\Omega \subset \mathbb{R}^2$, $\rho$ is the fluid density, $p$ is the effective mean pressure, $\mu$ is the molecular dynamic viscosity, $\mu_t$ is the eddy viscosity, and $\mathbf{g}$ is gravitational acceleration. The pressure $p$ absorbs the isotropic turbulent normal stress $\frac{2}{3}\rho k$. Both benchmarks in this study are isothermal with constant fluid properties, so the energy equation is not solved. Under the Boussinesq eddy-viscosity hypothesis, the deviatoric Reynolds stress tensor $\boldsymbol{\tau}^R = -\rho \overline{\mathbf{u}' \otimes \mathbf{u}'} + \frac{2}{3}\rho k \mathbf{I}$ is modeled linearly in terms of the mean strain-rate tensor $\mathbf{S}$:
\begin{equation}
    \boldsymbol{\tau}^R = 2 \mu_t \mathbf{S}, \qquad \mathbf{S} = \frac{1}{2}\!\left(\nabla \mathbf{u} + \nabla \mathbf{u}^{\mathsf{T}}\right).
    \label{eq:boussinesq}
\end{equation}
The high-fidelity reference and the baseline coarse-mesh code differ in how they close the scalar field $\mu_t(\mathbf{x}, t)$.

\subsection{High-Fidelity Reference: $k$--$\tau$ RANS Closure in Nek5000}
\label{sec:nek_ktau}

The reference data come from the spectral-element code Nek5000~\cite{Merzari01092020} with the two-equation $k$--$\tau$ RANS model~\cite{Tomboulides2025KTau,Nek5000RANSTheory}. The $k$--$\tau$ model is derived from the Wilcox $k$--$\omega$ model by introducing the turbulence timescale $\tau \equiv 1/(\beta^* \omega) = k/\varepsilon$. Both $k$ and $\tau$ vanish smoothly at no-slip walls, where $k = 0$ and $\tau = 0$ on $\partial\Omega_w$, so the model avoids the singular wall boundary condition of $\omega$. This property suits high-order spectral-element discretizations of curved-wall domains~\cite{Tomboulides2025KTau}.

In the $k$--$\tau$ formulation, the reference eddy viscosity is defined as
\begin{equation}
    \mu_t^* = \rho k \tau,
    \label{eq:mut_ktau}
\end{equation}
where the turbulent kinetic energy $k$ and turbulence timescale $\tau$ evolve according to two coupled advection--diffusion--reaction transport PDEs~\cite{Tomboulides2025KTau,Nek5000RANSTheory}:
\begin{align}
    \rho \frac{\partial k}{\partial t} + \rho (\mathbf{u}\cdot\nabla)k &= \nabla \cdot \!\left[\left(\mu + \frac{\mu_t}{\sigma_k}\right)\nabla k\right] + \mathcal{P}_k - \frac{\rho k}{\tau}, \label{eq:k_pde} \\
    \rho \frac{\partial \tau}{\partial t} + \rho (\mathbf{u}\cdot\nabla)\tau &= \nabla \cdot \!\left[\left(\mu + \frac{\mu_t}{\sigma_\tau}\right)\nabla \tau\right] - \gamma \frac{\tau}{k}\mathcal{P}_k + \rho \beta_1 - 2\frac{\mu}{k}\nabla k \cdot \nabla \tau \nonumber \\
    &\quad + \frac{2}{\tau}\!\left(\mu + \frac{\mu_t}{\sigma_\tau}\right)\nabla\tau\cdot\nabla\tau + \rho \sigma_d \frac{\tau^2}{k}\max\!\left(\nabla k \cdot \nabla \omega^*,\, 0\right), \label{eq:tau_pde}
\end{align}
in which $\mathcal{P}_k = \boldsymbol{\tau}^R : \nabla\mathbf{u} = 2\mu_t \|\mathbf{S}\|_F^2$ is the shear production of turbulent kinetic energy, $\omega^* = 1/(\beta^* \max(\tau, \tau_{\min}))$, and $(\sigma_k, \sigma_\tau, \gamma, \beta_1, \beta^*, \sigma_d)$ are standard closure constants~\cite{Tomboulides2025KTau}. Solving Eqs.~\eqref{eq:k_pde}--\eqref{eq:tau_pde} accurately requires resolving the steep near-wall gradients of $\nabla k$ and $\nabla \tau$ with fine boundary-layer elements at $y^+ \sim \mathcal{O}(1)$, which is infeasible on the coarse meshes used by system-level codes.

In this high-to-low study, the two-dimensional axisymmetric Nek5000 $k$--$\tau$ RANS solution is the assumed reference. Neither RANS nor the axisymmetric assumption fully captures the three-dimensional turbulence of the MSFR core. The comparisons therefore benchmark the closure against this reference and do not validate it against experiments or scale-resolving simulations. The aim is to show that the neural closure recovers on the coarse SAM mesh the two-equation RANS behavior that the mixing-length model loses.

\subsection{Coarse-Mesh Target Solver: Zero-Equation Closure in SAM}
\label{sec:sam_zeroeq}

In SAM's multidimensional porous-medium and open-fluid finite-element module~\cite{hu2025sam,zou2020development}, the conservation equations~\eqref{eq:continuity} and~\eqref{eq:momentum} are discretized using first-order Lagrange ($Q_1$) quadrilateral elements stabilized with the Streamline-Upwind/Petrov--Galerkin and Pressure-Stabilizing/Petrov--Galerkin (SUPG/PSPG) formulations~\cite{zou2020development,hu2025sam_manual}. Time integration uses the implicit backward Euler scheme with adaptive time steps. At each step, Newton's method in the MOOSE framework~\cite{lindsay2021automatic,hu2025sam} solves the nonlinear system with a finite-difference Jacobian, computed with graph coloring, and a direct LU factorization.

By default, SAM closes the eddy viscosity with Prandtl's algebraic zero-equation mixing-length model~\cite{bradshaw1974possible,hu2025sam_manual}:
\begin{equation}
    \mu_t^{\mathrm{zero\text{-}eq}} = \rho \ell_m^2 \sqrt{2\,\mathbf{S} : \mathbf{S}} = \sqrt{2}\,\rho \ell_m^2 \|\mathbf{S}\|_F,
    \label{eq:zero_eq}
\end{equation}
where $\|\mathbf{S}\|_F = \sqrt{\mathrm{tr}(\mathbf{S}^{\mathsf{T}}\mathbf{S})}$ is the Frobenius norm of the strain-rate tensor and $\ell_m$ is a capped Van Driest mixing length defined solely in terms of the minimum distance $d_w(\mathbf{x})$ to the nearest solid wall:
\begin{equation}
    \ell_m(d_w) = \min(\kappa d_w,\, \ell_{m,\max}), \qquad \ell_{m,\max} = 0.09\,\delta_b,
    \label{eq:mixing_length}
\end{equation}
with von K\'arm\'an constant $\kappa = 0.41$ and user-prescribed boundary-layer thickness $\delta_b$. Because Eq.~\eqref{eq:zero_eq} depends only on the pointwise strain-rate magnitude $\|\mathbf{S}(\mathbf{x})\|_F$ and geometric wall distance $d_w(\mathbf{x})$, it enforces $\mu_t^{\mathrm{zero\text{-}eq}} = 0$ wherever $\|\mathbf{S}\|_F = 0$, for example in vortex cores and at shear-layer inflection points. It also cannot represent the convective and diffusive transport of turbulent kinetic energy from high-shear inlet jets into central recirculating cavities.

\FloatBarrier
\section{Neural-Network Closure and Iterative Training}
\label{sec:methodology}

\begin{figure}[tbp]
    \centering
    \includegraphics[width=1\linewidth]{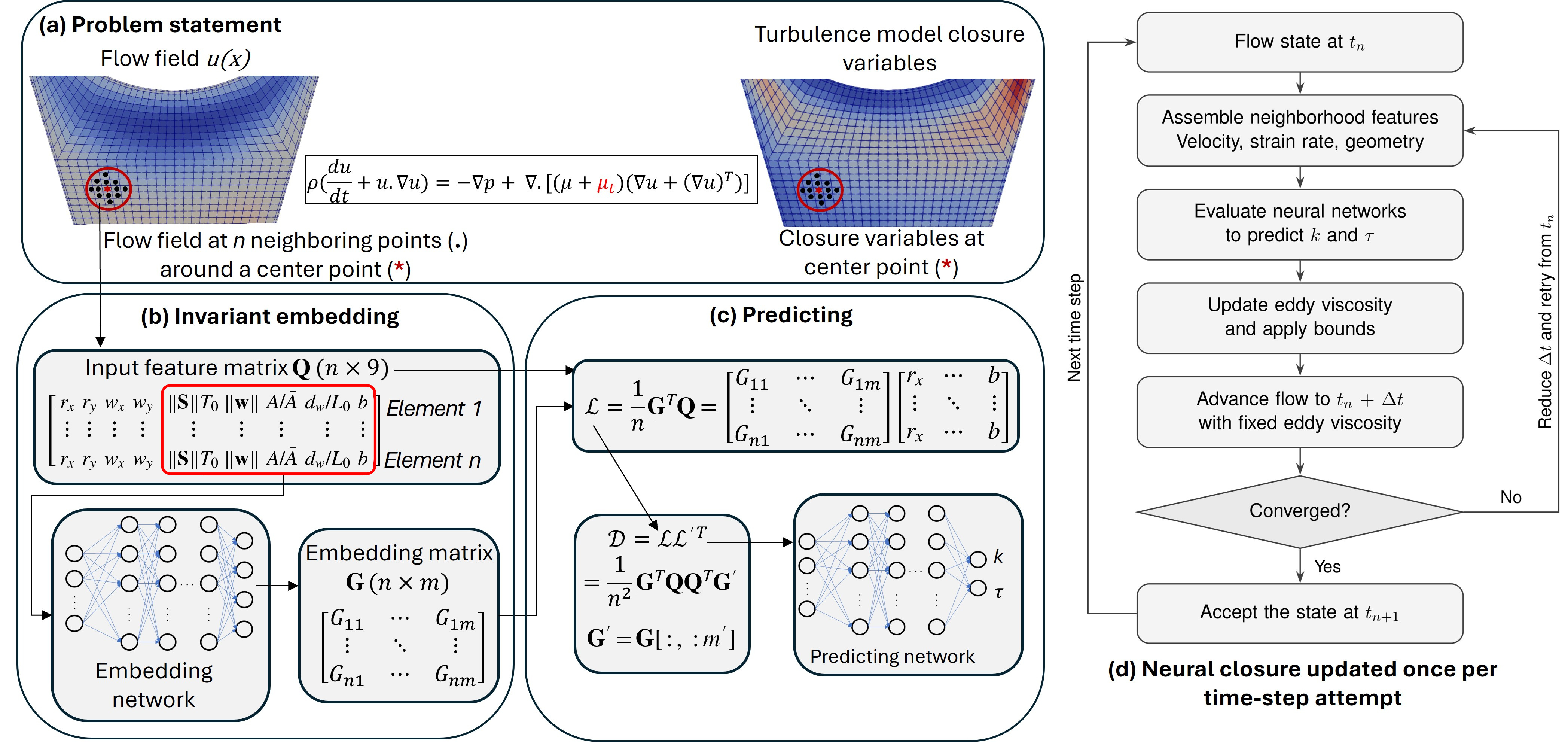}
\caption{Overview of the closure and its coupling with SAM. (a)~The eddy viscosity $\mu_t$ at each target element ($\ast$) is inferred from the flow state on its $n$ neighboring elements ($\bullet$). (b)~A shared-weight embedding network maps the scalar features of each row of $\mathbf{Q}\in\mathbb{R}^{n\times 9}$ to the embedding matrix $\mathbf{G}$. (c)~The invariant descriptor $\mathcal{D}$ (Eq.~\eqref{eq:D_descriptor}) and $Re$ are mapped to $\widehat{k}$ and $\widehat{\tau}$, giving $\widehat{\mu}_t=\rho\widehat{k}\widehat{\tau}\ge 0$. (d)~$\widehat{\mu}_t$ is evaluated once per time step from the accepted state, capped at 30~Pa\,s, and held fixed during the Newton iterations; a failed step is retried with a smaller $\Delta t$.}
    \label{fig:methodology}
\end{figure}

The goal is $k$--$\tau$ closure fidelity in SAM on a coarse mesh without solving the stiff transport PDEs~\eqref{eq:k_pde}--\eqref{eq:tau_pde}. We build a neural closure that takes its inputs from neighboring elements of the SAM mesh and train it iteratively with SAM. Figure~\ref{fig:methodology} illustrates the overall two-stage framework. Stage~1 trains offline: Nek5000 $k$--$\tau$ fields are projected onto the coarse mesh to initialize the neural closure. Stage~2 trains iteratively: the closure runs inside SAM's Newton solver and is retrained on the element flow states from all coupled runs. The regression targets stay fixed at the projected Nek5000 turbulence fields.

\subsection{Neighborhood and Input Features}
\label{sec:feature_engineering}

In a non-equilibrium recirculating flow on a coarse mesh, the eddy viscosity $\mu_t(\mathbf{x}_0, t)$ of a target element $0$ with centroid $\mathbf{x}_0$ depends on its local strain rate. It also depends on the velocity, shear, and mesh resolution of the neighboring elements. In SAM~\cite{lindsay2021automatic,zou2020development,hu2025sam}, individual elements span large physical volumes, and near-wall regions have only a few element layers. Rather than sampling points from a continuous or fine-grid point set~\cite{ZHOU2021113927,ZHOU2022114211,Xu2022PDEFree}, we formulate the closure directly on the finite elements of the computational mesh. For each target element $0$, we define its neighborhood $\mathcal{N}_0$ as the target element together with its edge-adjacent elements and their edge-adjacent elements, which form a two-layer topological neighborhood:
\begin{equation}
    \mathcal{N}_0 = \{0\} \cup \mathcal{E}(0) \cup \bigcup_{b \in \mathcal{E}(0)} \mathcal{E}(b), \qquad n = |\mathcal{N}_0|,
    \label{eq:neighborhood_def}
\end{equation}
where element indices are taken from $\{1, 2, \dots, N_e\}$, the label $0$ denotes the target element, which is itself included in $\mathcal{N}_0$, and $\mathcal{E}(a)$ is the set of elements that share an edge with element $a$. The neighborhood size $n$ varies with the local mesh topology and the proximity to walls or the symmetry axis. On the coarse SAM mesh, $n$ ranges from 6 to 14, with a mean of $12.2$. A topological neighborhood is used instead of a fixed Euclidean radius because it adapts automatically to the local element size, requires no mesh-dependent radius parameter, and is built directly from the element connectivity available in SAM at runtime.

Let $L_0$, $U_0$, and $T_0 = L_0 / U_0$ denote the characteristic length, velocity, and convective time scales of the flow system. Each neighbor $a \in \mathcal{N}_0$ has a centroid $\mathbf{x}_a$, velocity $\mathbf{u}_a$, strain-rate tensor $\mathbf{S}_a$, area $A_a$, minimum wall distance $d_{w,a}$, and wall-adjacency indicator $b_a \in \{0, 1\}$. From these we build a dimensionless feature row $\mathbf{q}_a^{\mathsf{T}} \in \mathbb{R}^{1\times 9}$ with two relative vectors and five scalar features:
\begin{equation}
    \mathbf{q}_a^{\mathsf{T}} = \big[\,\mathbf{r}_a^{\mathsf{T}},\; \mathbf{w}_a^{\mathsf{T}},\; \mathbf{c}_a^{\mathsf{T}}\,\big] \in \mathbb{R}^{1\times 9},
    \label{eq:feature_row}
\end{equation}
where the dimensionless relative position vector $\mathbf{r}_a \in \mathbb{R}^2$ and relative velocity vector $\mathbf{w}_a \in \mathbb{R}^2$ are measured with respect to the target element $0$:
\begin{equation}
    \mathbf{r}_a = \frac{\mathbf{x}_a - \mathbf{x}_0}{L_0}, \qquad
    \mathbf{w}_a = \frac{\mathbf{u}_a - \mathbf{u}_0}{U_0}.
    \label{eq:rel_vectors}
\end{equation}
The scalar features $\mathbf{c}_a \in \mathbb{R}^5$ follow from a fixed componentwise logarithmic map $\psi_c: \mathbb{R}_+^4 \times \{0, 1\} \to \mathbb{R}^5$ applied to five dimensionless invariants of neighbor $a$:
\begin{equation}
    \mathbf{c}_a = \psi_c\!\left(\|\mathbf{S}_a\|_F T_0,\; \|\mathbf{w}_a\|_2,\; \frac{A_a}{\bar{A}},\; \frac{d_{w,a}}{L_0},\; b_a\right) \in \mathbb{R}^5,
    \label{eq:feat_c}
\end{equation}
where $\bar{A} = \frac{1}{N_e}\sum_{e=1}^{N_e} A_e$ is the domain-mean element area and $b_a = 1$ if element $a$ is adjacent to a physical wall and $b_a = 0$ if it lies in the fluid interior. The normalized area $A_a/\bar{A}$ and wall indicator $b_a$ of every neighbor tell the closure about mesh grading and under-resolved near-wall elements. The logarithmic map $\psi_c$ compresses the multi-decade ranges of the strain rate $\|\mathbf{S}_a\|_F T_0$ and relative speed $\|\mathbf{w}_a\|_2$ into $\mathcal{O}(1)$ intervals that suit network training~\cite{yin2020feature}. Stacking the $n$ neighbor rows defines the feature matrix $\mathbf{Q} \in \mathbb{R}^{n\times 9}$ and its scalar submatrix $\mathbf{C} \in \mathbb{R}^{n\times 5}$:
\begin{equation}
    \mathbf{Q} = \begin{bmatrix} \mathbf{q}_1^{\mathsf{T}} \\ \vdots \\ \mathbf{q}_n^{\mathsf{T}} \end{bmatrix} \in \mathbb{R}^{n\times 9}, \qquad
    \mathbf{C} = \begin{bmatrix} \mathbf{c}_1^{\mathsf{T}} \\ \vdots \\ \mathbf{c}_n^{\mathsf{T}} \end{bmatrix} \in \mathbb{R}^{n\times 5}.
    \label{eq:Q_block}
\end{equation}

\subsection{Invariant Embedding and Prediction Architecture}
\label{sec:nn_architecture}

Because the number of neighboring elements $n$ varies across unstructured mesh locations and the indexing order of rows in $\mathbf{Q}$ is arbitrary, a fixed-input multilayer perceptron (MLP) cannot be applied directly to $\mathbf{Q}$. Instead, we employ a two-stage invariant architecture consisting of a shared-weight \textit{embedding network} $\phi_{\mathrm{emb}}$ and a \textit{prediction network} $\phi_{\mathrm{pred}}$~\cite{zhu2019machine,ZHOU2022114211,Xu2022PDEFree}.

First, the embedding network $\phi_{\mathrm{emb}}: \mathbb{R}^5 \to \mathbb{R}^m$ acts identically and independently on each row $\mathbf{c}_a^{\mathsf{T}}$ of the scalar feature submatrix $\mathbf{C} \in \mathbb{R}^{n\times 5}$ to produce an $m$-channel embedding matrix $\mathbf{G} \in \mathbb{R}^{n\times m}$:
\begin{equation}
    \mathbf{G}_{a:} = \phi_{\mathrm{emb}}\!\left(\mathbf{c}_a^{\mathsf{T}};\, \Theta_{\mathrm{emb}}\right) \in \mathbb{R}^{1\times m}, \qquad a = 1, \dots, n,
    \label{eq:embedding_G}
\end{equation}
and we denote by $\mathbf{G}' \in \mathbb{R}^{n\times m'}$ ($m' < m$) the reduced embedding submatrix formed by the first $m'$ columns of $\mathbf{G}$.

Second, $\mathbf{G}$ and $\mathbf{G}'$ project the feature matrix $\mathbf{Q} \in \mathbb{R}^{n\times 9}$ into the neighborhood-averaged feature matrix $\mathcal{L} \in \mathbb{R}^{m\times 9}$ and the fixed-size invariant descriptor matrix $\mathcal{D} \in \mathbb{R}^{m\times m'}$:
\begin{equation}
    \mathcal{L} = \frac{1}{n}\mathbf{G}^{\mathsf{T}}\mathbf{Q} \in \mathbb{R}^{m\times 9}, \qquad
    \mathcal{D} = \frac{1}{n^2}\mathbf{G}^{\mathsf{T}}\mathbf{Q}\mathbf{Q}^{\mathsf{T}}\mathbf{G}' \in \mathbb{R}^{m\times m'}.
    \label{eq:D_descriptor}
\end{equation}
Regardless of the neighborhood cardinality $n$, $\mathcal{D}$ has fixed dimension $m \times m'$ and vectorizes to $\mathrm{vec}(\mathcal{D}) \in \mathbb{R}^{mm'}$.

The embedding network acts on one neighbor at a time. For each of its $m$ channels, it learns how strongly that neighbor contributes to the averages in Eq.~\eqref{eq:D_descriptor}, so the closure can emphasize, for example, wall-adjacent or high-shear neighbors. Using the scalar features $\mathbf{C}$ directly as these weights would tie each neighbor's contribution linearly to its feature values. Neighbors with near-zero features would then be ignored, and physically distinct neighborhoods could map to the same descriptor.

Third, the prediction network $\phi_{\mathrm{pred}}: \mathbb{R}^{mm'+1} \to \mathbb{R}^2$ maps the concatenation of $\mathrm{vec}(\mathcal{D})$ and the global operating Reynolds number $Re = \rho U_{\mathrm{in}} L_0 / \mu$ through a fully connected MLP to predict the dimensionless transformed turbulence vector $\widehat{\mathbf{z}} = (\widehat{z}_k, \widehat{z}_\tau)^{\mathsf{T}}$:
\begin{equation}
    \widehat{\mathbf{z}} = \begin{bmatrix} \widehat{z}_k \\ \widehat{z}_\tau \end{bmatrix} = \phi_{\mathrm{pred}}\!\left(\big[\mathrm{vec}(\mathcal{D})^{\mathsf{T}},\, Re\big]^{\mathsf{T}};\, \Theta_{\mathrm{pred}}\right).
    \label{eq:prediction_net}
\end{equation}
The dimensionless targets $\mathbf{z} = (z_k, z_\tau)^{\mathsf{T}}$ are defined by a smooth componentwise logarithmic bijection $\psi_z: \mathbb{R}_+^2 \to \mathbb{R}^2$. Its inverse returns positive $\widehat{k}$ and $\widehat{\tau}$ over several decades, and hence a non-negative eddy viscosity:
\begin{equation}
    \mathbf{z} = \psi_z\!\left(\frac{k}{U_0^2},\; \frac{\tau}{T_0}\right), \qquad
    (\widehat{k},\, \widehat{\tau}) = \psi_z^{-1}(\widehat{\mathbf{z}}) > 0, \qquad
    \widehat{\mu}_t = \rho \widehat{k}\widehat{\tau} \ge 0.
    \label{eq:reconstructed_mut}
\end{equation}

\subsection{Frame and Permutation Invariance}
\label{sec:invariance_proofs}

A constitutive closure must give the same prediction under a change of inertial frame, meaning translation, Galilean boost, rotation or reflection, and under any re-indexing of the neighbors~\cite{Speziale1998,Xu2022PDEFree,kaszuba2025implicit}. For any parameters $\Theta = (\Theta_{\mathrm{emb}}, \Theta_{\mathrm{pred}})$, the mapping $\mathcal{F}_\Theta: (\mathbf{Q}, Re) \mapsto (\widehat{k}, \widehat{\tau}, \widehat{\mu}_t)$ defined by Eqs.~\eqref{eq:feature_row}--\eqref{eq:reconstructed_mut} satisfies all four properties exactly. This follows by tracking the feature rows $\mathbf{q}_a^{\mathsf{T}} = [\mathbf{r}_a^{\mathsf{T}},\, \mathbf{w}_a^{\mathsf{T}},\, \mathbf{c}_a^{\mathsf{T}}]$ through the descriptor $\mathcal{D}$ of Eq.~\eqref{eq:D_descriptor}. Under a uniform spatial translation $\tilde{\mathbf{x}} = \mathbf{x} + \mathbf{x}_c$ ($\mathbf{x}_c \in \mathbb{R}^2$) and Galilean boost $\tilde{\mathbf{x}}(t) = \mathbf{x}(t) - \mathbf{u}_{\mathrm{b}} t$, $\tilde{\mathbf{u}}(\tilde{\mathbf{x}}, t) = \mathbf{u}(\mathbf{x}, t) - \mathbf{u}_{\mathrm{b}}$ ($\mathbf{u}_{\mathrm{b}} \in \mathbb{R}^2$), the uniform offsets $\mathbf{x}_c - \mathbf{u}_{\mathrm{b}} t$ and $-\mathbf{u}_{\mathrm{b}}$ cancel identically in the relative vectors of every neighbor $a \in \mathcal{N}_0$:
\begin{equation}
    \tilde{\mathbf{r}}_a = \frac{\tilde{\mathbf{x}}_a - \tilde{\mathbf{x}}_0}{L_0} = \mathbf{r}_a, \qquad
    \tilde{\mathbf{w}}_a = \frac{\tilde{\mathbf{u}}_a - \tilde{\mathbf{u}}_0}{U_0} = \mathbf{w}_a.
\end{equation}
Because the translation Jacobian is the identity and $\nabla \mathbf{u}_{\mathrm{b}} = \mathbf{0}$, the spatial gradient operator and strain-rate tensor are preserved ($\tilde{\nabla}\tilde{\mathbf{u}}_a = \nabla\mathbf{u}_a$ and $\tilde{\mathbf{S}}_a = \mathbf{S}_a$), as are $\|\tilde{\mathbf{w}}_a\|_2 = \|\mathbf{w}_a\|_2$, element areas $\tilde{A}_a = A_a$, wall distances $\tilde{d}_{w,a} = d_{w,a}$, and boundary indicators $b_a$. Consequently, every argument of $\psi_c$ in Eq.~\eqref{eq:feat_c} is invariant, so $\tilde{\mathbf{c}}_a = \mathbf{c}_a$ and $\tilde{\mathbf{Q}} = \mathbf{Q}$. Moreover, the Reynolds number $Re = \rho U_{\mathrm{in}} L_0 / \mu$ uses the inlet speed relative to the vessel walls, so $\mathcal{F}_\Theta(\tilde{\mathbf{Q}}, Re) = \mathcal{F}_\Theta(\mathbf{Q}, Re)$.

For orthogonal coordinate transformations $\tilde{\mathbf{x}} = \mathbf{R}\mathbf{x}$ and $\tilde{\mathbf{u}} = \mathbf{R}\mathbf{u}$ with any orthogonal matrix $\mathbf{R} \in O(2)$ satisfying $\mathbf{R}^{\mathsf{T}}\mathbf{R} = \mathbf{R}\mathbf{R}^{\mathsf{T}} = \mathbf{I}_2$ ($\det \mathbf{R} = \pm 1$), the relative vectors and strain-rate tensor transform objectively as
\begin{equation}
    \tilde{\mathbf{r}}_a = \mathbf{R}\mathbf{r}_a, \qquad
    \tilde{\mathbf{w}}_a = \mathbf{R}\mathbf{w}_a, \qquad
    \tilde{\mathbf{S}}_a = \mathbf{R}\mathbf{S}_a\mathbf{R}^{\mathsf{T}}.
\end{equation}
By the orthogonality of $\mathbf{R}$ and the cyclic property of the matrix trace, the relative speed and strain-rate Frobenius norm are invariant:
\begin{equation}
    \|\tilde{\mathbf{w}}_a\|_2 = \sqrt{\mathbf{w}_a^{\mathsf{T}}\mathbf{R}^{\mathsf{T}}\mathbf{R}\mathbf{w}_a} = \|\mathbf{w}_a\|_2, \qquad
    \|\tilde{\mathbf{S}}_a\|_F = \sqrt{\mathrm{tr}\!\left(\mathbf{R}\mathbf{S}_a^{\mathsf{T}}\mathbf{R}^{\mathsf{T}}\mathbf{R}\mathbf{S}_a\mathbf{R}^{\mathsf{T}}\right)} = \|\mathbf{S}_a\|_F.
\end{equation}
Since cell areas $A_a$ ($|\det \mathbf{R}| = 1$), Euclidean wall distances $d_{w,a}$, and boundary indicators $b_a$ are unchanged under $O(2)$, the scalar invariant vector satisfies $\tilde{\mathbf{c}}_a = \mathbf{c}_a$, or $\tilde{\mathbf{C}} = \mathbf{C}$. Because $\phi_{\mathrm{emb}}$ depends only on $\mathbf{C}$, the embedding matrices are invariant: $\tilde{\mathbf{G}} = \phi_{\mathrm{emb}}(\tilde{\mathbf{C}}) = \mathbf{G}$ and $\tilde{\mathbf{G}}' = \mathbf{G}'$. Meanwhile, each feature row of $\mathbf{Q}$ is right-multiplied by the block-diagonal orthogonal matrix $\mathbf{B}_R = \mathrm{diag}(\mathbf{R}^{\mathsf{T}},\, \mathbf{R}^{\mathsf{T}},\, \mathbf{I}_5) \in O(9)$:
\begin{equation}
    \tilde{\mathbf{q}}_a^{\mathsf{T}} = \big[\,\mathbf{r}_a^{\mathsf{T}}\mathbf{R}^{\mathsf{T}},\; \mathbf{w}_a^{\mathsf{T}}\mathbf{R}^{\mathsf{T}},\; \mathbf{c}_a^{\mathsf{T}}\,\big] = \mathbf{q}_a^{\mathsf{T}}\mathbf{B}_R \quad \Longrightarrow \quad \tilde{\mathbf{Q}} = \mathbf{Q}\mathbf{B}_R.
\end{equation}
Since $\mathbf{B}_R \mathbf{B}_R^{\mathsf{T}} = \mathrm{diag}(\mathbf{R}^{\mathsf{T}}\mathbf{R},\, \mathbf{R}^{\mathsf{T}}\mathbf{R},\, \mathbf{I}_5) = \mathbf{I}_9$, substituting $\tilde{\mathbf{Q}} = \mathbf{Q}\mathbf{B}_R$ and $\tilde{\mathbf{G}} = \mathbf{G}$ into Eq.~\eqref{eq:D_descriptor} yields
\begin{equation}
    \tilde{\mathcal{D}} = \frac{1}{n^2}\tilde{\mathbf{G}}^{\mathsf{T}}\tilde{\mathbf{Q}}\tilde{\mathbf{Q}}^{\mathsf{T}}\tilde{\mathbf{G}}'
    = \frac{1}{n^2}\mathbf{G}^{\mathsf{T}}\mathbf{Q}\underbrace{\mathbf{B}_R \mathbf{B}_R^{\mathsf{T}}}_{\mathbf{I}_9}\mathbf{Q}^{\mathsf{T}}\mathbf{G}'
    = \mathcal{D}.
    \label{eq:D_rot_proof}
\end{equation}
Thus every entry of $\mathcal{D}$ is an exact $O(2)$ invariant, and $\mathcal{F}_\Theta(\tilde{\mathbf{Q}}, Re) = \mathcal{F}_\Theta(\mathbf{Q}, Re)$.

Finally, under any neighbor-index permutation $\pi \in S_n$ with permutation matrix $\mathbf{P}_\pi \in \{0, 1\}^{n\times n}$ satisfying $\mathbf{P}_\pi^{\mathsf{T}}\mathbf{P}_\pi = \mathbf{I}_n$, reordering the neighboring elements gives $\tilde{\mathbf{Q}} = \mathbf{P}_\pi \mathbf{Q}$ and $\tilde{\mathbf{C}} = \mathbf{P}_\pi \mathbf{C}$. Because $\phi_{\mathrm{emb}}$ acts identically row-by-row on $\mathbf{C}$, it is permutation-equivariant: $\tilde{\mathbf{G}} = \phi_{\mathrm{emb}}(\mathbf{P}_\pi \mathbf{C}) = \mathbf{P}_\pi \mathbf{G}$ and $\tilde{\mathbf{G}}' = \mathbf{P}_\pi \mathbf{G}'$. Substituting into Eq.~\eqref{eq:D_descriptor} gives
\begin{equation}
    \tilde{\mathcal{L}} = \frac{1}{n}\tilde{\mathbf{G}}^{\mathsf{T}}\tilde{\mathbf{Q}}
    = \frac{1}{n}\mathbf{G}^{\mathsf{T}}\underbrace{\mathbf{P}_\pi^{\mathsf{T}}\mathbf{P}_\pi}_{\mathbf{I}_n}\mathbf{Q}
    = \mathcal{L}
    \quad \Longrightarrow \quad
    \tilde{\mathcal{D}} = \frac{1}{n}\tilde{\mathcal{L}}\tilde{\mathbf{Q}}^{\mathsf{T}}\tilde{\mathbf{G}}' = \mathcal{D},
\end{equation}
so the closure output is independent of neighbor indexing. In addition to exact permutation invariance under $S_n$, the factor $1/n$ in Eq.~\eqref{eq:D_descriptor} ensures that $\mathcal{L}_{kj} = \frac{1}{n}\sum_{a=1}^n G_{ak} Q_{aj}$ represents an empirical average over the $n$ elements of the neighborhood $\mathcal{N}_0$. Because the neighbor count $n$ differs between graded near-wall layers and the core, this $1/n$ normalization keeps $\mathcal{L}$ and $\mathcal{D}$ from scaling with $n$.

\subsection{Iterative Training}
\label{sec:iterative_theory}

The closure addresses the first two challenges of Section~\ref{sec:intro} by construction, and the iterative training addresses the third. First, $\widehat{\mu}_t = \rho\widehat{k}\widehat{\tau} \ge 0$ by Eq.~\eqref{eq:reconstructed_mut}, and $\mu + \widehat{\mu}_t$ enters the implicit diffusive operator of SAM's momentum equation~\eqref{eq:momentum} instead of an explicit source term~\cite{Wu2019IllConditioned}. Shan et al.~\cite{shan2025new} further showed that the spatial distribution and magnitude of the implicit eddy viscosity control the numerical stability of the coupled RANS system. As depicted in Fig.~\ref{fig:methodology}\textbf{d}, $\widehat{\mu}_t$ is evaluated from the accepted velocity state at the start of each time step and held fixed during the Newton iterations. This keeps the Jacobian free of network derivatives, and the added viscous term improves its diagonal dominance, as shown in Appendix~\ref{app:solver}. Second, the feature matrix $\mathbf{Q}$ resolves the multi-valued local strain-rate relation~\cite{Liu2021IterativeML,man2024nonunique} by combining the surrounding velocity and shear with wall distance, element area, and wall adjacency.

\begin{figure}[tbp]
    \centering
    \includegraphics[width=0.92\linewidth]{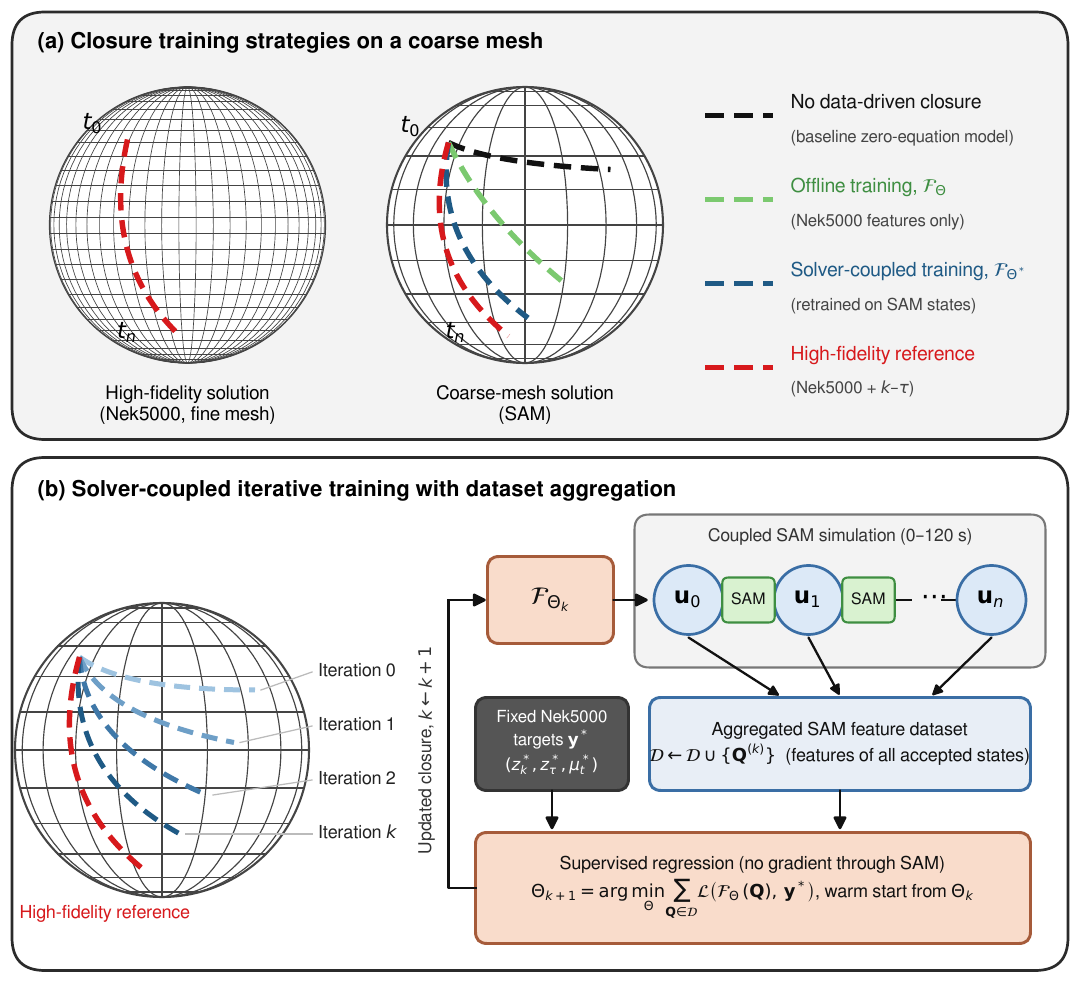}
    \caption{Iterative training. \textbf{(a)}~Training strategies on the coarse mesh: no neural closure, offline training on Nek5000 features, and retraining on SAM states. \textbf{(b)}~Iterative training: each coupled SAM run adds its feature states to the aggregated dataset $\mathcal{A}$, and the closure is retrained on $\mathcal{A}$ against the fixed Nek5000 targets without differentiating through the solver.}
    \label{fig:iteration}
\end{figure}

The third challenge, covariate shift, remains. The velocity and strain-rate fields resolved on SAM's coarse $Q_1$ mesh differ systematically from the projected Nek5000 fields. To remove this covariate shift~\cite{Liu2021IterativeML}, we use the iterative training depicted in Fig.~\ref{fig:iteration}. After Stage~1 initializes the network offline on projected Nek5000 snapshots, Stage~2 alternates between two steps. It runs coupled SAM simulations over the transient or steady-state horizon. It then retrains the network parameters $\Theta$ on the SAM feature matrices $\mathbf{Q}$, aggregated over all previous coupled runs, against the \textit{fixed} projected Nek5000 targets $(z_k^*, z_\tau^*, \mu_t^*)$. Network optimization minimizes, over mini-batches $\mathcal{B}$ of element-time samples,
\begin{equation}
    J(\Theta) = \frac{1}{|\mathcal{B}|}\sum_{a \in \mathcal{B}} \left[ w_k \left(\widehat{z}_{k,a} - z_{k,a}^*\right)^2 + w_\tau \left(\widehat{z}_{\tau,a} - z_{\tau,a}^*\right)^2 + w_\mu \frac{V_a}{\bar{V}} \frac{\left(\widehat{\mu}_{t,a} - \mu_{t,a}^*\right)^2}{\langle \mu_t^{*2} \rangle_V} \right],
    \label{eq:loss}
\end{equation}
where $V_a$ is the axisymmetric volume of the element of sample $a$, $\bar{V}$ is its mean over the training set, $\langle \mu_t^{*2} \rangle_V$ is the volume-weighted mean of $\mu_t^{*2}$ over the training set, and $(w_k, w_\tau, w_\mu) = (2.5, 1, 1)$, as listed in Table~\ref{tab:closure_hyperparams}. The penalty on the reconstructed eddy viscosity $\widehat{\mu}_t = \rho\widehat{k}\widehat{\tau}$ prevents opposite-signed errors in $\widehat{z}_k$ and $\widehat{z}_\tau$ from compounding in the momentum equation~\eqref{eq:momentum}. Retraining on the latest run makes the coupled errors oscillate between iterations, as shown in Section~\ref{sec:aposteriori_results}. We therefore retrain on all previous runs, whose feature states are kept in the aggregated dataset $\mathcal{A}$. Algorithm~\ref{alg:iterative_framework} summarizes the two training stages and the adaptive time stepping.

\begin{algorithm}[tbp]
\caption{Two-stage closure training: offline training (Stage~1) and iterative training (Stage~2).}
\label{alg:iterative_framework}
\small
\begin{algorithmic}[1]
\Require High-fidelity Nek5000 reference fields $(\mathbf{u}^*, k^*, \tau^*)$; coarse SAM mesh ($N_e$ elements).
\Ensure Trained closure parameters $\Theta$ and coupled coarse-mesh SAM predictions.
\Statex \textbf{Stage 1: Offline target projection and initial training}
\State Project Nek5000 reference fields onto the coarse SAM mesh to obtain fixed targets $(z_k^*, z_\tau^*, \mu_t^*)$ and baseline feature matrices $\mathbf{Q}$.
\State Train initial network parameters $\Theta$ on the training snapshots by minimizing the loss of Eq.~\eqref{eq:loss}, and export as a TorchScript module for C++ runtime inference in SAM.
\Statex \textbf{Stage 2: Iterative training}
\State Initialize the aggregated SAM feature dataset $\mathcal{A} \gets \emptyset$.
\For{each outer training iteration}
  \State Initialize coarse-mesh SAM simulation with the current TorchScript closure $\mathcal{F}_\Theta$.
  \While{$t < t_{\mathrm{end}}$}
    \State At the beginning of a time-step, assemble neighborhood features $\mathbf{Q}$ from the accepted SAM velocity state $\mathbf{u}$ and evaluate $\widehat{\mu}_t = \mathcal{F}_\Theta(\mathbf{Q}, Re)$ across all $N_e$ elements.
    \State Solve the nonlinear SUPG/PSPG finite-element system by Newton's method with a finite-difference (graph-colored) Jacobian and LU factorization, with $\mu + \widehat{\mu}_t$ embedded in the implicit diffusive operator of Eq.~\eqref{eq:momentum}.
    \State If Newton converges, accept the step and record the realized SAM features $\mathbf{Q}$; otherwise, reduce $\Delta t$ by a factor of $0.8$ and retry.
  \EndWhile
  \State Aggregate the realized SAM feature states of this coupled run: $\mathcal{A} \gets \mathcal{A} \cup \{\mathbf{Q}\}$.
  \State Warm-start retrain $\Theta$ on all of $\mathcal{A}$ against the fixed targets $(z_k^*, z_\tau^*, \mu_t^*)$, select the checkpoint on the held-out interior snapshots, and export the updated TorchScript module.
\EndFor
\end{algorithmic}
\end{algorithm}

\FloatBarrier
\section{MSFR Benchmark Configuration and Closure Implementation}
\label{sec:test_case}

\subsection{Axisymmetric Core Geometry and Multi-Fidelity Discretization}
\label{sec:geometry_mesh}

\begin{figure}[tbp]
    \centering
    \includegraphics[width=0.98\linewidth]{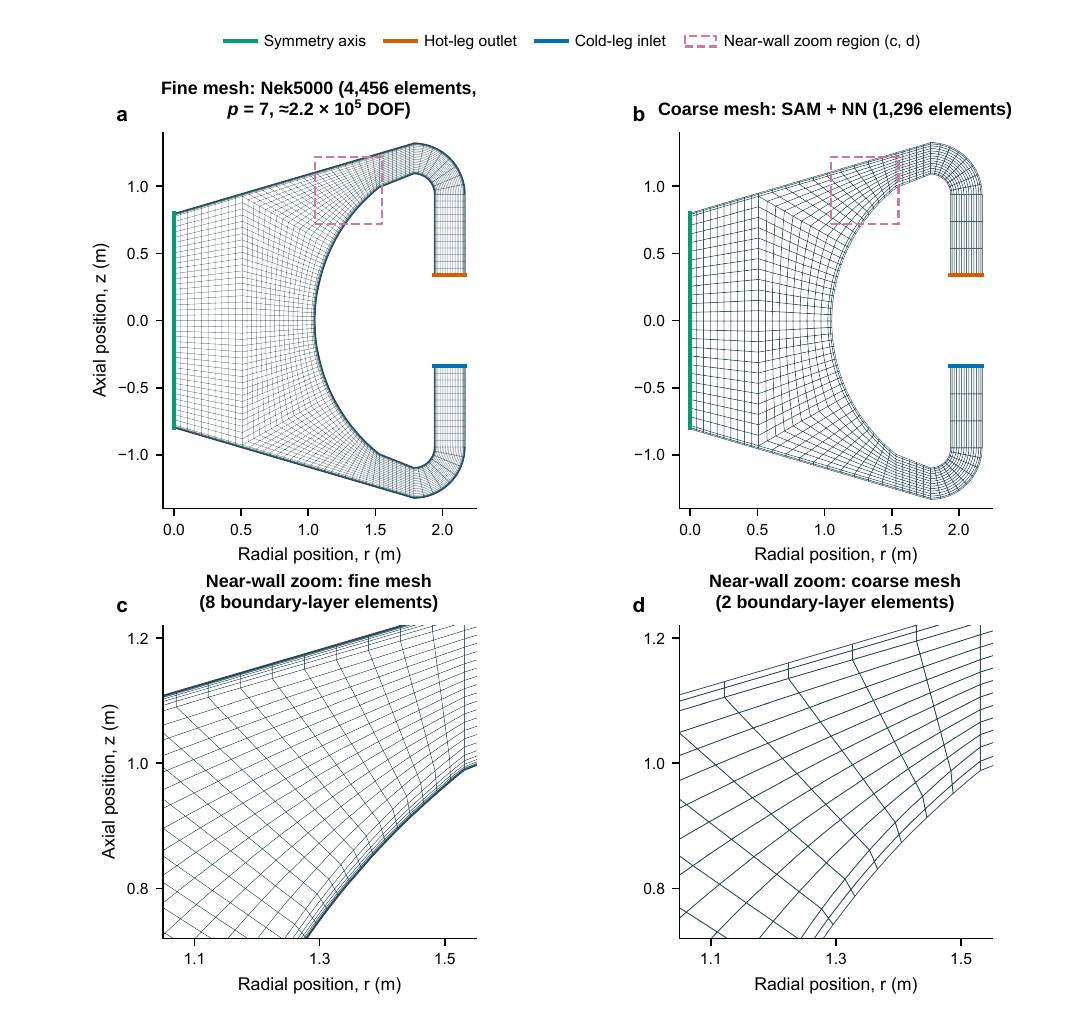}
    \caption{Computational meshes and boundary conditions for the axisymmetric MSFR cavity. \textbf{a},~Nek5000 spectral-element mesh ($4{,}456$ elements, $p = 7$). \textbf{b},~Coarse SAM mesh ($1{,}296$ elements). \textbf{c}, \textbf{d},~Near-wall views of the dashed region in Nek5000 (eight boundary-layer elements) and SAM (two).}
    \label{fig:mesh}
\end{figure}

We evaluate the closure on the two-dimensional axisymmetric meridional cross-section of the reference MSFR cavity~\cite{ROUCH2014449,fang2021fullcore,inlVTBmsfrNek5000}. As shown in Fig.~\ref{fig:mesh}, the computational domain spans $r \in [0.0,\, 2.17]$~m in the radial direction, with the symmetry axis at $r = 0$~m, and $z \in [-1.325,\, 1.325]$~m in the axial direction. Fuel salt enters the lower core leg through the cold-leg inlet at $z = -0.34$~m and flows upward through the central cavity. The cavity is bounded by a curved inner fertile-blanket/reflector shield wall and inclined upper and lower reflector walls. The salt exits through the hot-leg outlet at $z = +0.34$~m. Both openings span $r \in [1.935,\, 2.17]$~m.

Figure~\ref{fig:mesh} compares the Nek5000 reference mesh in Fig.~\ref{fig:mesh}\textbf{a} with the coarse SAM mesh used with the neural closure in Fig.~\ref{fig:mesh}\textbf{b}. Figures~\ref{fig:mesh}\textbf{c} and~\ref{fig:mesh}\textbf{d} show close-ups of the near-wall discretization along the upper reflector and curved inner shield wall. The Nek5000 mesh has $4{,}456$ spectral elements of polynomial order $p = 7$. With 8 Gauss--Lobatto--Legendre quadrature points per direction in each element, it has about $2.2 \times 10^5$ nodal degrees of freedom, and eight graded boundary-layer elements resolve the steep near-wall velocity and turbulence gradients. The SAM mesh has only $N_e = 1{,}296$ coarse elements, with two boundary-layer elements across the near-wall region. It has $70.9\%$ fewer elements than the Nek5000 mesh and more than two orders of magnitude fewer nodal degrees of freedom. For comparison with the mixing-length model, a coarse SAM mesh of $1{,}080$ bilinear ($Q_1$) elements is also evaluated, with $1{,}825$ elements for the steady states.

\subsection{Steady-State Operating Regimes}
\label{sec:test_case_steady}

The first benchmark campaign examines isothermal steady-state flow at six operating conditions spanning a factor of five in Reynolds number, $Re \in [4.52 \times 10^5,\, 2.26 \times 10^6]$. Table~\ref{tab:simulation_cases} summarizes the dimensionless parameters and thermophysical properties of the six steady-state cases. The steady-state Reynolds number $Re = \rho U_{\mathrm{in}} L_0/\mu$ uses the core diameter, $L_0 = 2.0$~m. The transient closure features instead use the cold-leg inlet width, $L_0 = 0.235$~m, as listed in Table~\ref{tab:closure_hyperparams}. Cases~1--5 train the steady-state closure. Case~6, at $Re = 4.52 \times 10^5$, just below the training range, is held out to test steady-state generalization.

\begin{table}[tbp]
\centering
\caption{Operating parameters and thermophysical properties for the six steady-state MSFR cases.}
\label{tab:simulation_cases}
\footnotesize
\setlength{\tabcolsep}{3.5pt}
\begin{tabular}{@{} l c c c c c c @{}}
\toprule
\textbf{Parameter} & \textbf{Case 1} & \textbf{Case 2} & \textbf{Case 3} & \textbf{Case 4} & \textbf{Case 5} & \textbf{Case 6 (Test)} \\
\midrule
Reynolds number, $Re$ & $2.26\times 10^6$ & $1.50\times 10^6$ & $1.13\times 10^6$ & $9.03\times 10^5$ & $5.64\times 10^5$ & $4.52\times 10^5$ \\
Inlet speed, $U_{\mathrm{in}}$ (m/s) & 1.25 & 1.25 & 1.25 & 1.25 & 1.25 & 1.25 \\
Length scale, $L_0$ (m) & 2.0 & 2.0 & 2.0 & 2.0 & 2.0 & 2.0 \\
Density, $\rho$ (kg/m$^3$) & 4125 & 4125 & 4125 & 4125 & 4125 & 4125 \\
Viscosity, $\mu$ (Pa\,s) & 0.0046 & 0.0069 & 0.0091 & 0.0114 & 0.0183 & 0.0228 \\
\bottomrule
\end{tabular}
\end{table}

\subsection{Transient Pump-Coastdown Benchmark}
\label{sec:test_case_transient}

The second benchmark campaign evaluates a 120~s pump-coastdown transient starting from a fully developed steady-state flow field. During the transient, the cold-leg inlet velocity profile decays exponentially according to
\begin{align}
    U_{\mathrm{pk}}(t) &= U_\infty + \left(U_{\mathrm{pk}}(0) - U_\infty\right)\exp\!\left(-\frac{\ln 2}{t_{1/2}}\,t\right), \label{eq:transient_upk} \\
    u_{\mathrm{in}}(r, t) &= -C_Q\,\eta(r)\,U_{\mathrm{pk}}(t), \qquad C_Q = 1.0273064, \label{eq:transient_inlet}
\end{align}
where $U_{\mathrm{pk}}(0) = 1.25$~m/s is the initial peak inlet velocity, $U_\infty = 0.30$~m/s is the asymptotic circulation speed, $t_{1/2} = 20$~s is the excess-velocity half-life, and $\eta(r)$ is the inlet velocity profile prescribed in the Nek5000 simulation, normalized by its centerline value. $U_{\mathrm{pk}}(t)$ is evaluated exactly at every SAM time step, as in the Nek5000 simulation. The quadrature correction factor $C_Q$ ensures exact conservation of the integrated volumetric inflow rate when $\eta(r)$ is represented on the coarse SAM inlet boundary. No-slip boundary conditions are enforced along all solid core walls, axial symmetry is imposed at $r = 0$~m, and a constant outlet pressure of $161$~kPa is prescribed at the hot-leg exit. As in the Nek5000 simulation, the flow is isothermal and no body force is applied. Table~\ref{tab:transient_settings} lists the physical properties and simulation settings for the transient benchmark.

\begin{table}[tbp]
\centering
\caption{Thermophysical properties and boundary-condition parameters for the 120~s transient flow-decay benchmark.}
\label{tab:transient_settings}
\footnotesize
\begin{tabular}{l l}
\toprule
\textbf{Quantity} & \textbf{Value} \\
\midrule
Fluid density, $\rho$ & $4306.71$~kg/m$^3$ \\
Molecular dynamic viscosity, $\mu$ & $0.00894$~Pa\,s \\
Molecular kinematic viscosity, $\nu$ & $2.08\times 10^{-6}$~m$^2$/s \\
Reference CFD turbulence closure & Two-equation $k$--$\tau$ RANS \\
Initial peak inlet velocity, $U_{\mathrm{pk}}(0)$ & $1.25$~m/s \\
Asymptotic peak inlet velocity, $U_\infty$ & $0.30$~m/s \\
Excess-velocity decay half-life, $t_{1/2}$ & $20$~s \\
Transient duration, $t_{\mathrm{end}}$ & $120$~s \\
\bottomrule
\end{tabular}
\end{table}

To assess temporal interpolation and extrapolation across the 120~s transient, the Nek5000 trajectory is sampled at one-second intervals and projected onto the $1{,}296$ elements of the coarse SAM mesh. The training set $\mathcal{T}_{\mathrm{tr}}$ has 56 complete snapshots in the window $t \in [11,\, 69]$~s. The snapshots at $t = 20$, $40$, and $60$~s in this window, $3{,}888$ element-time samples, are held out to test \textit{interior temporal interpolation}. The snapshots at $t \in [2,\, 10]$~s, $11{,}664$ samples, test \textit{early temporal extrapolation} during the high-speed initial deceleration. Those at $t \in [70,\, 120]$~s, $66{,}096$ samples, test \textit{late temporal extrapolation} in the low-speed asymptotic circulation. Each snapshot is assigned as a complete spatial field to either training or testing, preventing spatial data leakage between neighboring elements at the same time step.

\subsection{Feature Scaling, Network Hyperparameters, and Training Configuration}
\label{sec:model_hyperparams}

Table~\ref{tab:closure_hyperparams} lists the reference scales, the normalization maps $\psi_c$ and $\psi_z$, and the network hyperparameters used for the MSFR benchmark. For the transient benchmark, the scales in Eqs.~\eqref{eq:rel_vectors}--\eqref{eq:feat_c} are the cold-leg inlet width $L_0 = 0.235$~m, the initial peak inlet velocity $U_0 = 1.25$~m/s, and $T_0 = L_0 / U_0 = 0.188$~s. At MSFR operating conditions, $Re \sim 10^6$, $\mu \approx 0.009$~Pa\,s, and $\mu_t \sim 5$--$20$~Pa\,s, so the eddy-to-molecular viscosity ratio reaches $\mu_t / \mu \sim \mathcal{O}(10^3)$. The eddy viscosity therefore dominates the viscous term of Eq.~\eqref{eq:momentum}, so it is treated implicitly there. Appendix~\ref{app:solver} shows its effect on the Newton solver.

The maps $\psi_c$ and $\psi_z$ are logarithmic, with the forms listed in Table~\ref{tab:closure_hyperparams}. They compress the multi-decade ranges of strain rate, relative speed, $k$, and $\tau$ across the MSFR jet, boundary layer, and recirculation core into $\mathcal{O}(1)$ intervals. The wall distance is capped at $L_0$, and $A_a/\bar{A}$ and $b_a$ enter unscaled. The element strain-rate magnitude $\|\mathbf{S}_a\|_F$ is computed from the full two-dimensional velocity-gradient tensor at the element quadrature points, both at runtime in SAM and in the offline projection of the Nek5000 fields.

The embedding network $\phi_{\mathrm{emb}}$ is a three-layer MLP with $m = 64$ channels, of which the first $m' = 8$ form $\mathbf{G}'$. The descriptor $\mathcal{D}$ thus has $64 \times 8 = 512$ entries. The prediction network $\phi_{\mathrm{pred}}$ is a four-layer MLP that takes $\mathrm{vec}(\mathcal{D})$ together with $Re(t)$, a $513$-dimensional input. Both networks use leaky rectified linear unit (leaky ReLU) activations, and Table~\ref{tab:closure_hyperparams} lists the layer widths. The single-trajectory offline study omits $Re$, giving 512 inputs.

\begin{table}[tbp]
\centering
\caption{Case-specific normalization maps and neural-network architecture hyperparameters for the MSFR closure implementation.}
\label{tab:closure_hyperparams}
\footnotesize
\begin{tabular}{l l}
\toprule
\textbf{Parameter / Component} & \textbf{MSFR Implementation Specification} \\
\midrule
Characteristic scales $(L_0, U_0, T_0)$ & $0.235$~m ($2.0$~m steady-state), $1.25$~m/s, $0.188$~s \\
Strain-rate normalization in $\psi_c$, Eq.~\eqref{eq:feat_c} & $\frac{1}{2}\ln\max(\|\mathbf{S}_a\|_F T_0,\, 10^{-12}) + 2$ \\
Relative-speed normalization in $\psi_c$, Eq.~\eqref{eq:feat_c} & $\frac{1}{6}\ln(1 + 100\max(\|\mathbf{w}_a\|_2,\, 10^{-8}))$ \\
Wall-distance normalization in $\psi_c$, Eq.~\eqref{eq:feat_c} & $\min(d_{w,a}/L_0,\, 1)$ \\
Target $k$ transform in $\psi_z$, Eq.~\eqref{eq:reconstructed_mut} & $z_k = \frac{1}{2}\ln(1 + 1250\,k/U_0^2)$ \\
Target $\tau$ transform in $\psi_z$, Eq.~\eqref{eq:reconstructed_mut} & $z_\tau = \frac{1}{4}[\ln\max(\tau/T_0,\, 10^{-12}) + 7]$ \\
Embedding MLP $\phi_{\mathrm{emb}}$ widths & $5 \to 32 \to 64 \to 64$ (leaky ReLU activations) \\
Embedding channels $(m, m')$ & $m = 64$, $m' = 8$ ($\mathcal{D} \in \mathbb{R}^{64\times 8}$, $\mathrm{vec}(\mathcal{D}) \in \mathbb{R}^{512}$) \\
Prediction MLP $\phi_{\mathrm{pred}}$ widths & $513 \to 128 \to 64 \to 8 \to 2$ (leaky ReLU activations) \\
Neighborhood $\mathcal{N}_0$, Eq.~\eqref{eq:neighborhood_def} & Two-layer edge adjacency, $n = 6$--$14$ (mean $12.2$) \\
Loss weights $(w_k, w_\tau, w_\mu)$, Eq.~\eqref{eq:loss} & $(2.5,\, 1,\, 1)$ \\
Optimizer & Adam, batch size 512, gradient-norm clipping at 5 \\
Learning rate & $10^{-3}$ (Stage~1), $10^{-4}$ (Stage~2), cosine decay to $10\%$ \\
Epochs & $\le 400$ (Stage~1), $\le 150$ per outer iteration \\
Checkpoint selection & Lowest $\mu_t$ error on held-out snapshots $t = 20, 40, 60$~s \\
\bottomrule
\end{tabular}
\end{table}

\FloatBarrier
\section{Results and Discussion}
\label{sec:results}

\subsection{A Priori Closure Evaluation on Held-Out Nek5000 Trajectories}
\label{sec:nek_apriori}

Before coupling the closure with SAM, we evaluate its \textit{a priori} accuracy on the 63 held-out Nek5000 snapshots, $81{,}648$ element-time samples in total, using fixed weights trained only on the 56 Nek5000 training snapshots. The Nek5000 velocity is projected onto the coarse SAM mesh, so that the network inputs are computed in the same way as at runtime in SAM. For any temporal subset of element-time samples $\mathcal{S}$, we quantify prediction accuracy using the pooled volume-weighted relative $L_2$ error and root-mean-square error (RMSE):
\begin{equation}
    E_q(\mathcal{S}) = \left[\frac{\sum_{a \in \mathcal{S}} V_a \left(\widehat{q}_a - q_a^*\right)^2}{\sum_{a \in \mathcal{S}} V_a \left(q_a^*\right)^2}\right]^{1/2}, \qquad
    \mathrm{RMSE}_q(\mathcal{S}) = \left[\frac{1}{|\mathcal{S}|}\sum_{a \in \mathcal{S}} \left(\widehat{q}_a - q_a^*\right)^2\right]^{1/2},
    \label{eq:apriori_metrics}
\end{equation}
where $q \in \{k, \tau, \mu_t\}$ denotes the physical turbulence variable reconstructed via Eq.~\eqref{eq:reconstructed_mut}, $q^*$ is the Nek5000 reference value, and $V_a$ is the axisymmetric volume of the element of sample $a$.

\begin{figure}[tbp]
    \centering
    \includegraphics[width=0.98\linewidth]{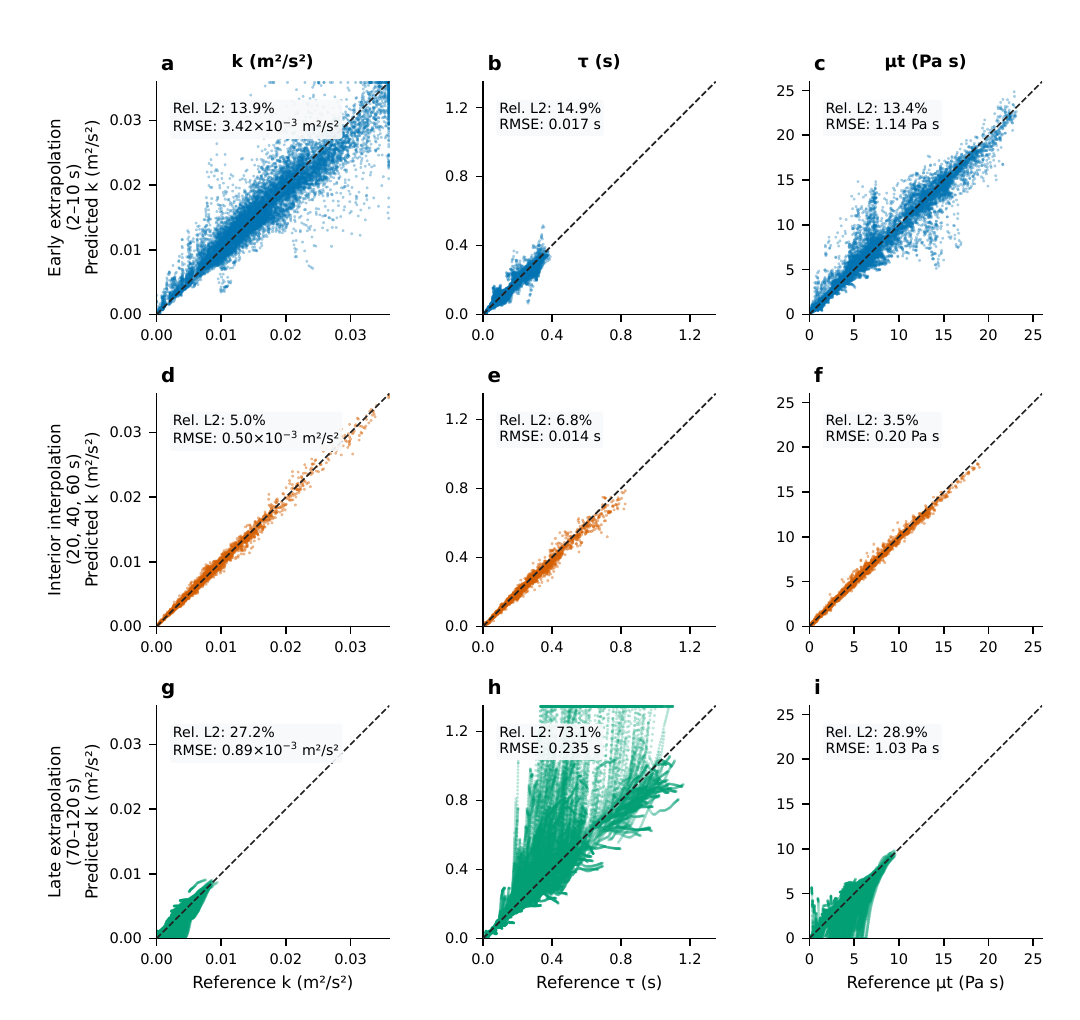}
    \caption{\textit{A priori} parity plots of the network predictions against the Nek5000 $k$--$\tau$ reference. Columns: $k$, $\tau$, and $\mu_t$. Rows: early extrapolation ($2$--$10$~s; \textbf{a}--\textbf{c}), interior interpolation ($20, 40, 60$~s; \textbf{d}--\textbf{f}), and late extrapolation ($70$--$120$~s; \textbf{g}--\textbf{i}). Dashed lines indicate 1:1 agreement.}
    \label{fig:nek_apriori_scatter}
\end{figure}

\begin{figure}[tbp]
    \centering
    \includegraphics[width=0.98\linewidth]{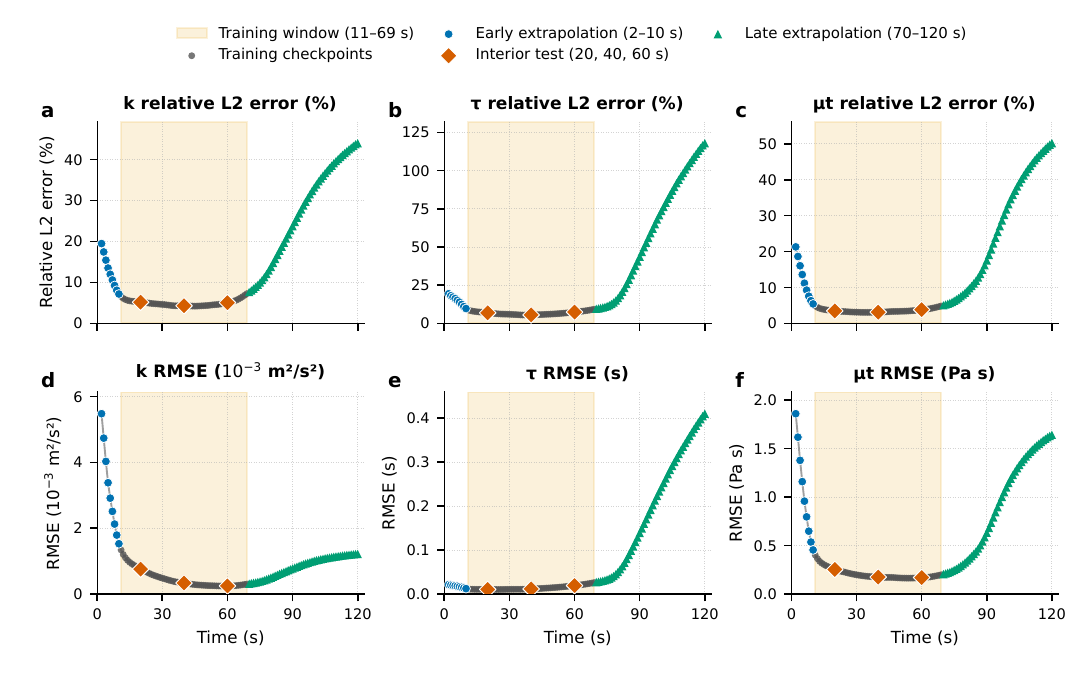}
    \caption{Temporal evolution of \textit{a priori} prediction errors across $t \in [2,\, 120]$~s. \textbf{a}--\textbf{c}, Relative $L_2$ errors for $k$ (\textbf{a}), $\tau$ (\textbf{b}), and $\mu_t$ (\textbf{c}). \textbf{d}--\textbf{f}, Physical-unit RMSE for $k$ (\textbf{d}), $\tau$ (\textbf{e}), and $\mu_t$ (\textbf{f}). The shaded amber band denotes the $11$--$69$~s training window, within which $t = 20, 40, 60$~s are held out for interior interpolation testing.}
    \label{fig:nek_apriori_errors}
\end{figure}

Figure~\ref{fig:nek_apriori_scatter} shows the \textit{a priori} parity plots for $k$, $\tau$, and $\mu_t$ in the three held-out temporal regimes. Figure~\ref{fig:nek_apriori_errors} tracks the instantaneous relative $L_2$ error and physical-unit RMSE at all 119 checkpoints from $t = 2$~s to $120$~s. In the interior interpolation regime at $t = 20$, $40$, and $60$~s, shown in Figs.~\ref{fig:nek_apriori_scatter}\textbf{d}--\textbf{f}, the pooled relative $L_2$ errors are $5.0\%$ for $k$, $6.8\%$ for $\tau$, and $3.5\%$ for $\mu_t$, with $\mathrm{RMSE}_{\mu_t} = 0.20$~Pa\,s.

In the early extrapolation regime at $t = 2$--$10$~s, shown in Figs.~\ref{fig:nek_apriori_scatter}\textbf{a}--\textbf{c}, the core flow experiences rapid initial deceleration from $U_{\mathrm{pk}}(0) = 1.25$~m/s, generating higher strain rates and turbulent kinetic energy than any state inside the $11$--$69$~s training window. Consequently, the network underpredicts the upper tail of $k$ at $t = 2$~s, with $E_k = 19.5\%$ and $\mathrm{RMSE}_{\mu_t} = 1.86$~Pa\,s. The error decays monotonically as $t$ approaches the start of the training window at $11$~s, as shown in Figs.~\ref{fig:nek_apriori_errors}\textbf{a} and~\ref{fig:nek_apriori_errors}\textbf{d}. The pooled early-extrapolation eddy-viscosity relative $L_2$ error is $13.4\%$, with $\mathrm{RMSE}_{\mu_t} = 1.14$~Pa\,s.

In the late extrapolation regime at $t = 70$--$120$~s, shown in Figs.~\ref{fig:nek_apriori_scatter}\textbf{g}--\textbf{i} and~\ref{fig:nek_apriori_errors}, the inlet velocity falls below the range covered by the training window, and the prediction error grows with time. The relative $L_2$ error in $\mu_t$ rises from $5.0\%$ at $t = 70$~s to $50.2\%$ at $t = 120$~s, with a pooled late-extrapolation error of $28.9\%$, as shown in Fig.~\ref{fig:nek_apriori_errors}\textbf{c}. Part of this rise comes from the decay of the reference norm $\|q^*(t)\|_2$ in the denominator of the relative error as the flow slows down. However, the physical-unit RMSE in $\mu_t$ also increases, from $0.21$ to $1.64$~Pa\,s with a pooled value of $1.03$~Pa\,s, as shown in Fig.~\ref{fig:nek_apriori_errors}\textbf{f}. The RMSE of $k$ stays below $1.3\times 10^{-3}$~m$^2$/s$^2$, as shown in Fig.~\ref{fig:nek_apriori_errors}\textbf{d}. The turbulence timescale is the least robust quantity in this regime, with a pooled $E_\tau = 73.1\%$, as Fig.~\ref{fig:nek_apriori_scatter}\textbf{h} shows. The late-time error is therefore a genuine extrapolation error and not only an effect of the decaying reference norm. Reporting both the relative $L_2$ error and the physical RMSE separates the two effects.

\subsection{A Posteriori Coupled Evaluation in SAM}
\label{sec:aposteriori_results}

\begin{figure}[tbp]
    \centering
    \includegraphics[width=0.82\linewidth]{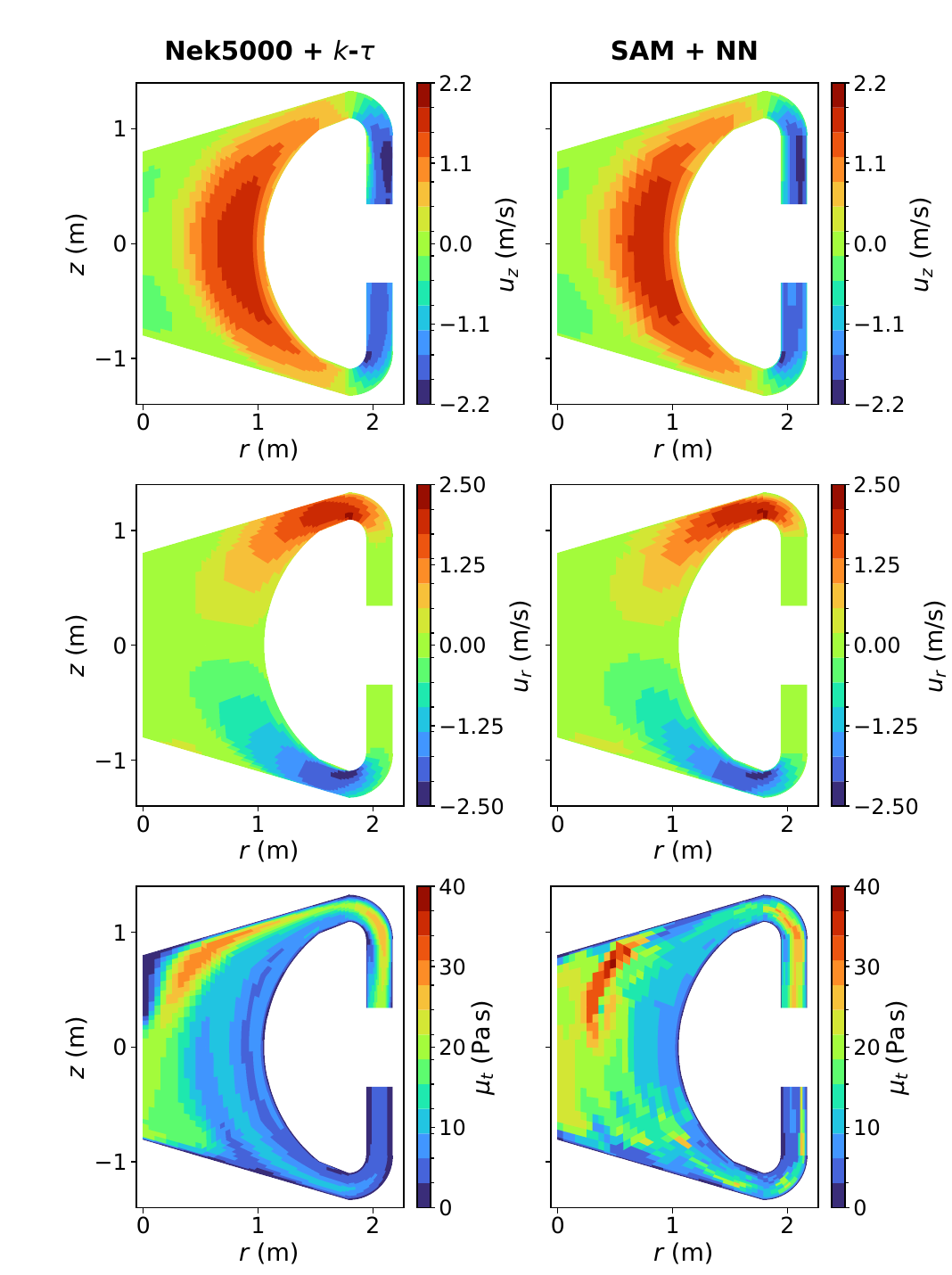}
    \caption{Steady-state \textit{a posteriori} comparison for held-out Case~6 ($Re = 4.52\times 10^5$): Nek5000 + $k$--$\tau$ (left) and SAM + NN (right). Rows: axial velocity $u_z$ (labeled $u_x$), negative radial velocity $-u_r$ (labeled $u_y$), and eddy viscosity $\mu_t$. Panel axes $x$ and $y$ correspond to $r$ and $z$. The symmetry axis is at the left edge and the cold-leg inlet at the bottom, as in all contour figures.}
    \label{fig:steady_state_results}
\end{figure}

We next evaluate the neural closure \textit{a posteriori}, embedded in SAM after the iterative training of Section~\ref{sec:iterative_theory}. Figure~\ref{fig:steady_state_results} compares the steady-state axial and radial velocity components ($u_z$ and $u_r$) and eddy-viscosity ($\mu_t$) fields on held-out Case~6 at $Re = 4.52 \times 10^5$. It shows the Nek5000 $k$--$\tau$ reference and SAM coupled with the neural closure, labeled SAM + NN in the figures and tables.

As shown in Fig.~\ref{fig:steady_state_results}, SAM coupled with the neural closure reproduces the single recirculation cell and the high-$\mu_t$ core plume of the Nek5000 $k$--$\tau$ reference on the coarse mesh. In the plume, $\mu_t \approx 16$--$25$~Pa\,s.

For the 120~s pump-coastdown transient, SAM states are available exactly at each one-second Nek5000 checkpoint, $t \in \{2, 3, \dots, 120\}$~s. At each checkpoint we compute the volume-weighted relative $L_2$ error on the $N_e = 1{,}296$ coarse mesh, along with its time average:
\begin{equation}
    E_q(t) = \left[\frac{\sum_{e=1}^{N_e} V_e \left\| q_e(t) - q_e^*(t) \right\|_2^2}{\sum_{e=1}^{N_e} V_e \left\| q_e^*(t) \right\|_2^2}\right]^{1/2}, \qquad
    \overline{E}_q = \frac{1}{119}\sum_{t=2}^{120} E_q(t),
    \label{eq:transient_error}
\end{equation}
where $q \in \{\mathbf{u}, \mu_t\}$ denotes the SAM velocity vector or eddy viscosity, $q^*$ is the projected Nek5000 reference field, and $V_e$ is the axisymmetric volume of coarse element $e$.

\begin{figure}[p]
    \centering
    \includegraphics[width=0.92\linewidth]{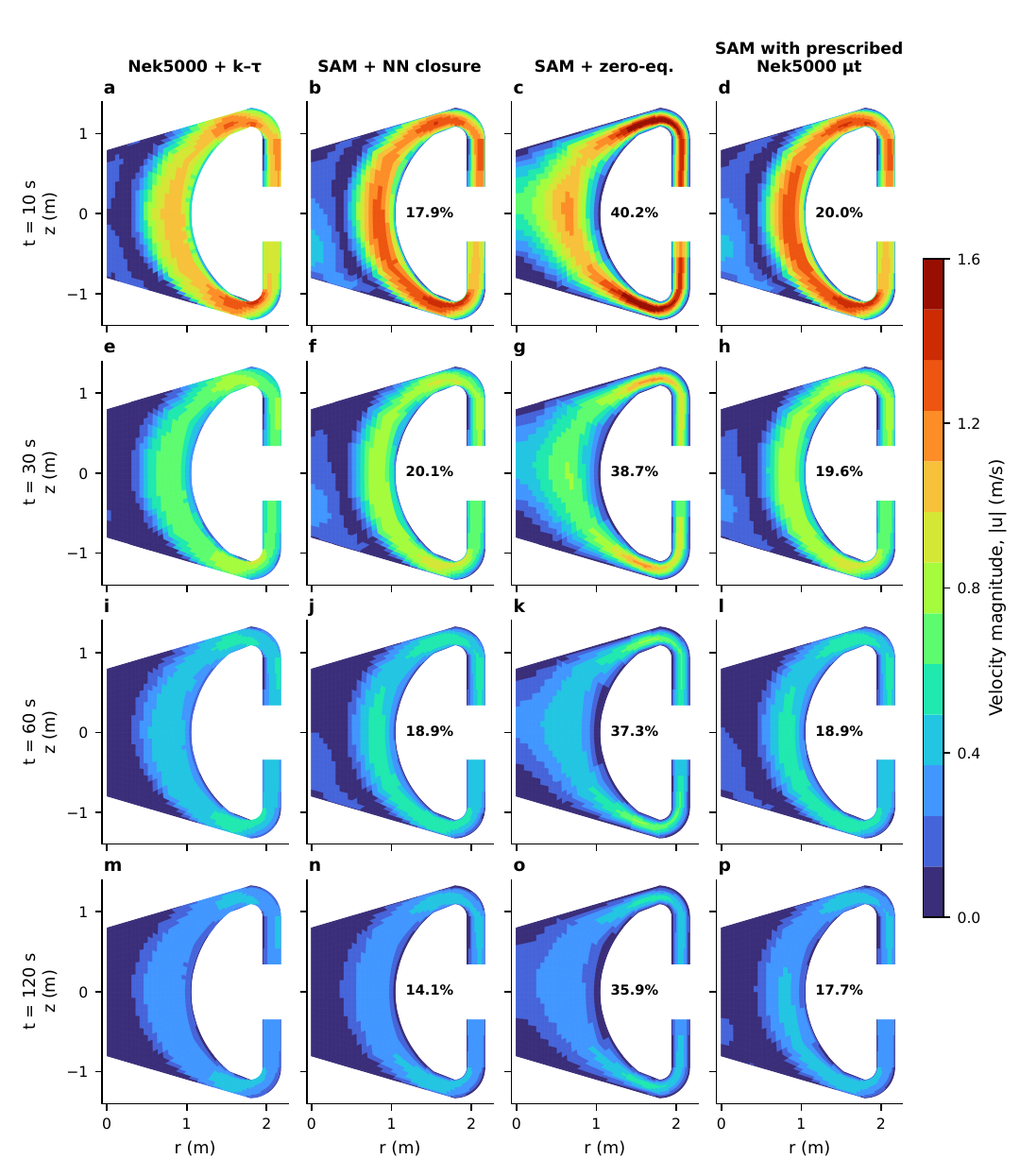}
    \caption{Velocity magnitude $|\mathbf{u}|$ at $t = 10, 30, 60,$ and $120$~s (rows). Columns: projected Nek5000 reference, SAM + NN, SAM + mixing length, and SAM with prescribed Nek5000 $\mu_t$; panels \textbf{a}--\textbf{p} are ordered row by row. In-panel labels give the relative $L_2$ velocity error.}
    \label{fig:velocity_transient}
\end{figure}

\begin{figure}[p]
    \centering
    \includegraphics[width=0.72\linewidth]{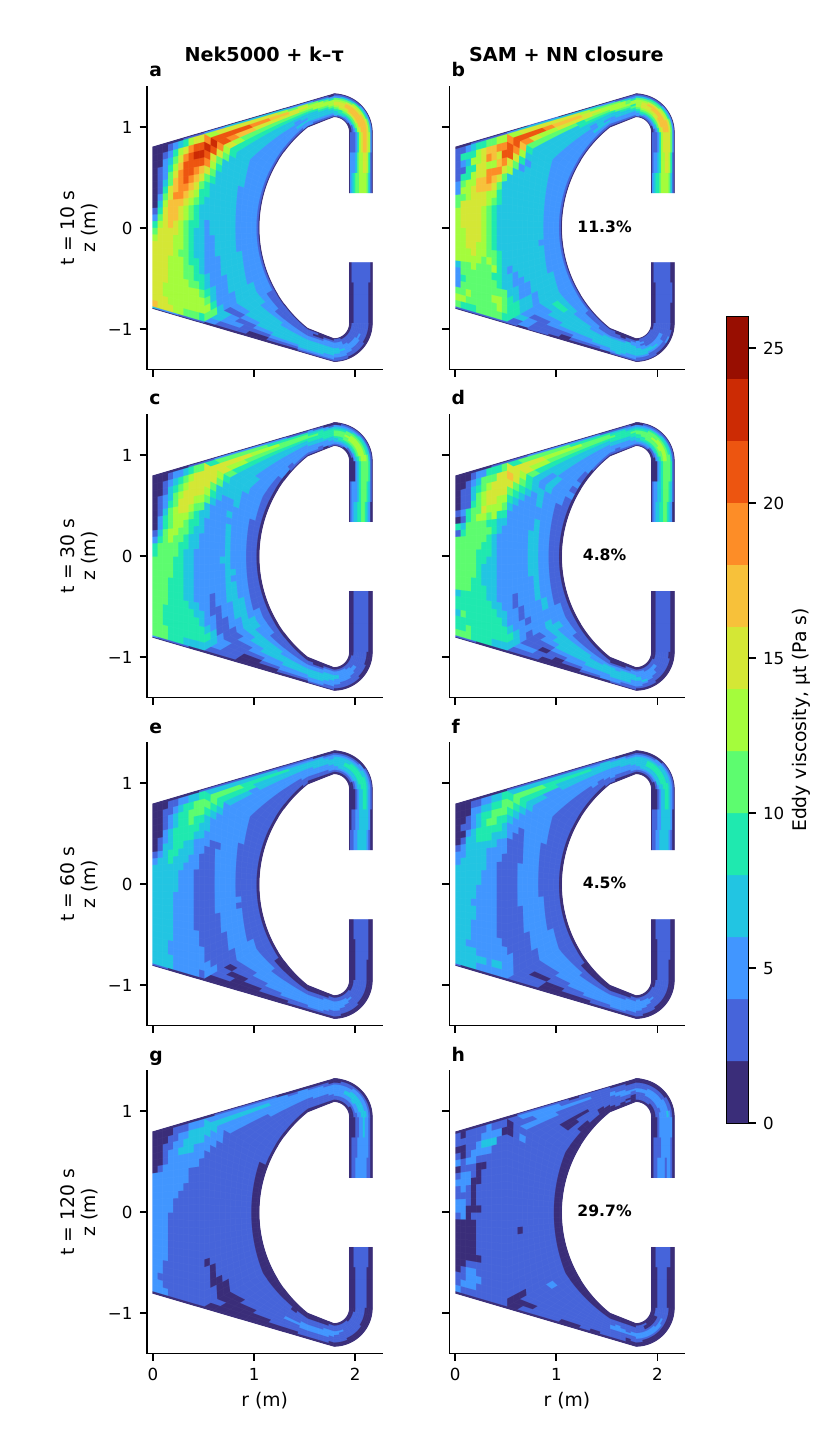}
    \caption{\textit{A posteriori} transient eddy-viscosity fields $\mu_t$ at $t = 10, 30, 60,$ and $120$~s (rows) comparing the projected Nek5000 + $k$--$\tau$ reference (\textbf{a}, \textbf{c}, \textbf{e}, \textbf{g}) and SAM + NN (\textbf{b}, \textbf{d}, \textbf{f}, \textbf{h}) on the 1,296-element coarse mesh. In-panel labels give the volume-weighted relative $L_2$ eddy-viscosity error.}
    \label{fig:viscosity_transient}
\end{figure}

Figures~\ref{fig:velocity_transient} and~\ref{fig:viscosity_transient} compare the spatial evolution of the velocity magnitude $|\mathbf{u}|$ and eddy viscosity $\mu_t$ at four snapshots: $t = 10$~s in early extrapolation, $t = 30$ and $60$~s in the training window, and $t = 120$~s at the end of late extrapolation. As shown in Figs.~\ref{fig:velocity_transient}\textbf{c}, \textbf{g}, \textbf{k}, and \textbf{o}, the mixing-length model in SAM underpredicts turbulent diffusion in the core interior. It concentrates the flow in a thin, fast jet along the outer cavity wall and leaves a weak, diffuse core flow, with a time-averaged velocity error of $37.6\%$. Figures~\ref{fig:velocity_transient}\textbf{b}, \textbf{f}, \textbf{j}, \textbf{n} and~\ref{fig:viscosity_transient}\textbf{b}, \textbf{d}, \textbf{f}, \textbf{h} show SAM coupled with the neural closure. It tracks the contraction of the high-speed stream along the curved inner wall, with a time-averaged velocity error of $17.7\%$. It also tracks the decay of the core eddy-viscosity plume from $\mu_t \approx 15$--$24$~Pa\,s at $t = 10$~s to $\mu_t \approx 3$--$5$~Pa\,s at $t = 120$~s. The last column of Fig.~\ref{fig:velocity_transient} shows SAM with prescribed Nek5000 $\mu_t$ in place of the network. Its velocity field is close to that of SAM with the neural closure, with a time-averaged error of $18.8\%$ against $17.7\%$. Table~\ref{tab:order_mesh} identifies the source of this shared error.

\begin{figure}[tbp]
    \centering
    \includegraphics[width=0.88\linewidth]{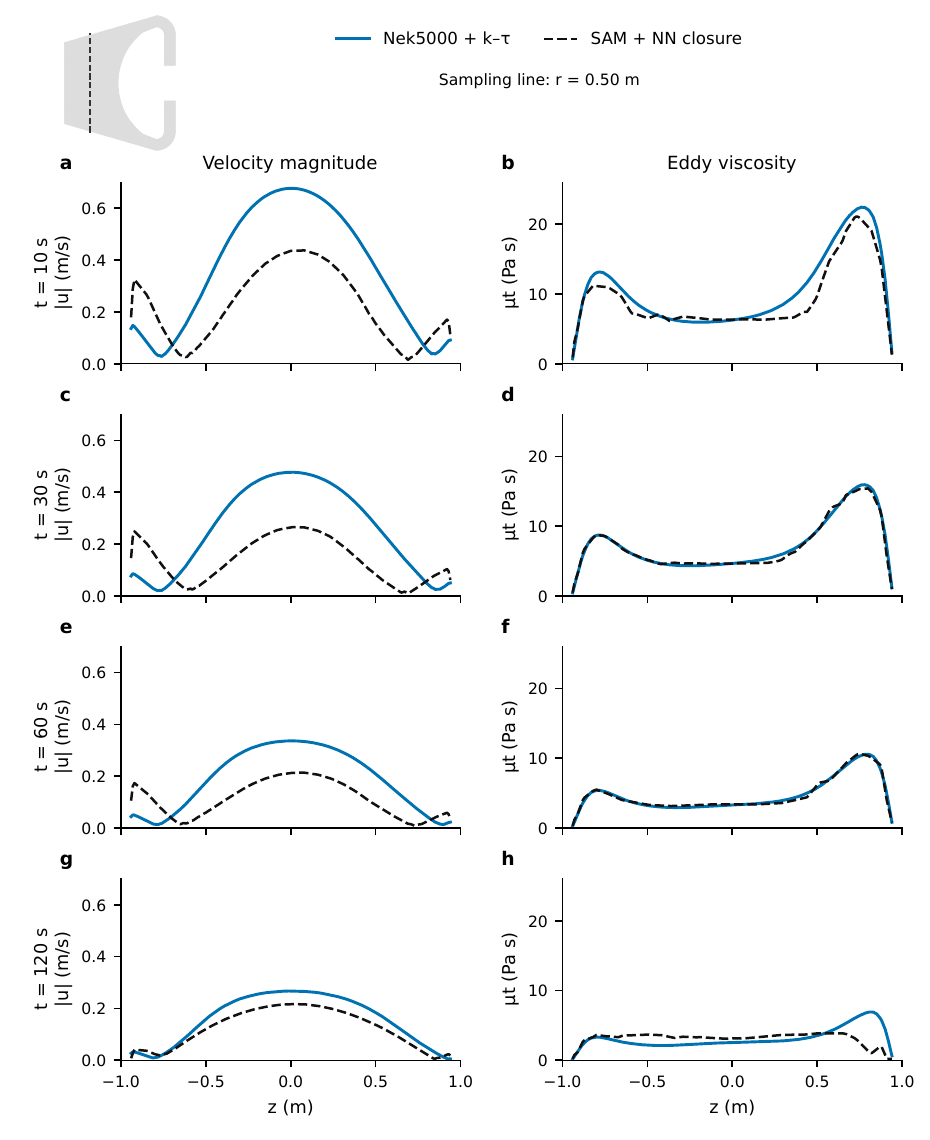}
    \caption{One-dimensional transient profiles of velocity magnitude $|\mathbf{u}|$ (left column) and eddy viscosity $\mu_t$ (right column) along the cross-core sampling line $r = 0.50$~m (indicated in the top-left inset) at $t = 10, 30, 60,$ and $120$~s.}
    \label{fig:transient_profiles}
\end{figure}

Figure~\ref{fig:transient_profiles} examines the one-dimensional profiles of $|\mathbf{u}|$ and $\mu_t$ along the interior sampling line $r = 0.50$~m at $t = 10, 30, 60,$ and $120$~s. Along this line, the eddy-viscosity profiles of SAM with the neural closure agree closely with the Nek5000 reference at $t = 30$ and $60$~s and fall below it near $z \approx -0.8$~m at $t = 10$~s. At $t = 120$~s, in the late extrapolation window, the near-wall peak at $z \approx 0.8$~m is missed. The velocity profiles have the correct shape. However, SAM with the neural closure underpredicts the peak core speed on this line by about $35$--$45\%$ at $t = 10$--$60$~s, for example $0.44$ versus $0.68$~m/s at $t = 10$~s, and by about $20\%$ at $t = 120$~s. SAM with prescribed Nek5000 $\mu_t$ has a similar velocity error, $18.8\%$ against $17.7\%$, so this difference does not originate in the closure. Table~\ref{tab:order_mesh} traces it to the coarse mesh.

\begin{table}[tbp]
\centering
\caption{Volume-weighted relative $L_2$ errors of SAM + NN relative to the projected Nek5000 reference across the 120~s transient, evaluated via Eq.~\eqref{eq:transient_error}.}
\label{tab:transient_errors}
\footnotesize
\setlength{\tabcolsep}{5pt}
\begin{tabular}{@{} l r r @{}}
\toprule
\textbf{Checkpoint / Temporal Statistic} & \textbf{Velocity Error $E_{\mathbf{u}}$ (\%)} & \textbf{Viscosity Error $E_{\mu_t}$ (\%)} \\
\midrule
$t = 10$~s (early extrapolation) & 17.88 & 11.28 \\
$t = 30$~s (training window) & 20.09 & 4.85 \\
$t = 60$~s (interior held-out checkpoint) & 18.90 & 4.51 \\
$t = 120$~s (end of late extrapolation) & 14.08 & 29.71 \\
\midrule
\textbf{Time-averaged error $\overline{E}_q$ ($2$--$120$~s)} & \textbf{17.73} & \textbf{12.02} \\
\bottomrule
\end{tabular}
\end{table}

\begin{figure}[tbp]
    \centering
    \includegraphics[width=0.98\linewidth]{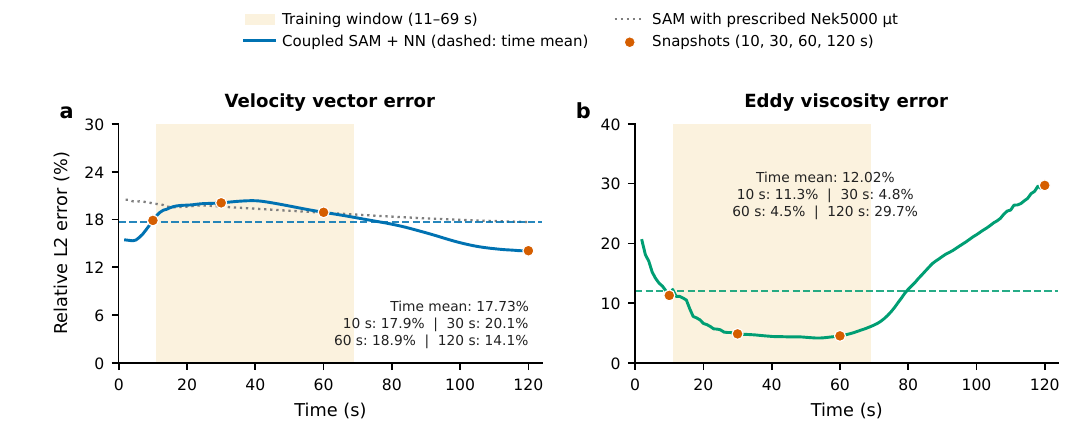}
    \caption{Temporal evolution of \textit{a posteriori} volume-weighted relative $L_2$ errors for SAM + NN across $t \in [2,\, 120]$~s. \textbf{a}, Velocity error $E_{\mathbf{u}}(t)$; the dotted line is SAM with prescribed Nek5000 $\mu_t$. \textbf{b}, Eddy-viscosity relative error $E_{\mu_t}(t)$. Dashed horizontal lines denote the time means ($17.73\%$ and $12.02\%$), and orange markers highlight the four snapshot times displayed in Figs.~\ref{fig:velocity_transient}--\ref{fig:transient_profiles}.}
    \label{fig:transient_errors}
\end{figure}

Figure~\ref{fig:transient_errors} and Table~\ref{tab:transient_errors} show the time history of the \textit{a posteriori} errors $E_{\mathbf{u}}(t)$ and $E_{\mu_t}(t)$ over the full 120~s transient. Between $t = 10$ and $60$~s, $E_{\mathbf{u}}(t)$ follows the error of SAM with prescribed Nek5000 $\mu_t$, shown as the dotted line in Fig.~\ref{fig:transient_errors}\textbf{a} with a time average of $18.8\%$. After $t \approx 65$~s, it falls below that curve, giving $\overline{E}_{\mathbf{u}} = 17.73\%$. The eddy-viscosity error is smallest inside the training window, at $4.85\%$ for $t = 30$~s and $4.51\%$ for $t = 60$~s. It grows in the late extrapolation window to $29.71\%$ at $t = 120$~s, following the \textit{a priori} trend of Section~\ref{sec:nek_apriori}, with a time mean of $\overline{E}_{\mu_t} = 12.02\%$. The coupled simulation completes all $169$ adaptive implicit time steps without a failed nonlinear solve. Appendix~\ref{app:solver} analyzes the effect of the closure on solver convergence.

\begin{figure}[tbp]
    \centering
    \includegraphics[width=0.99\linewidth]{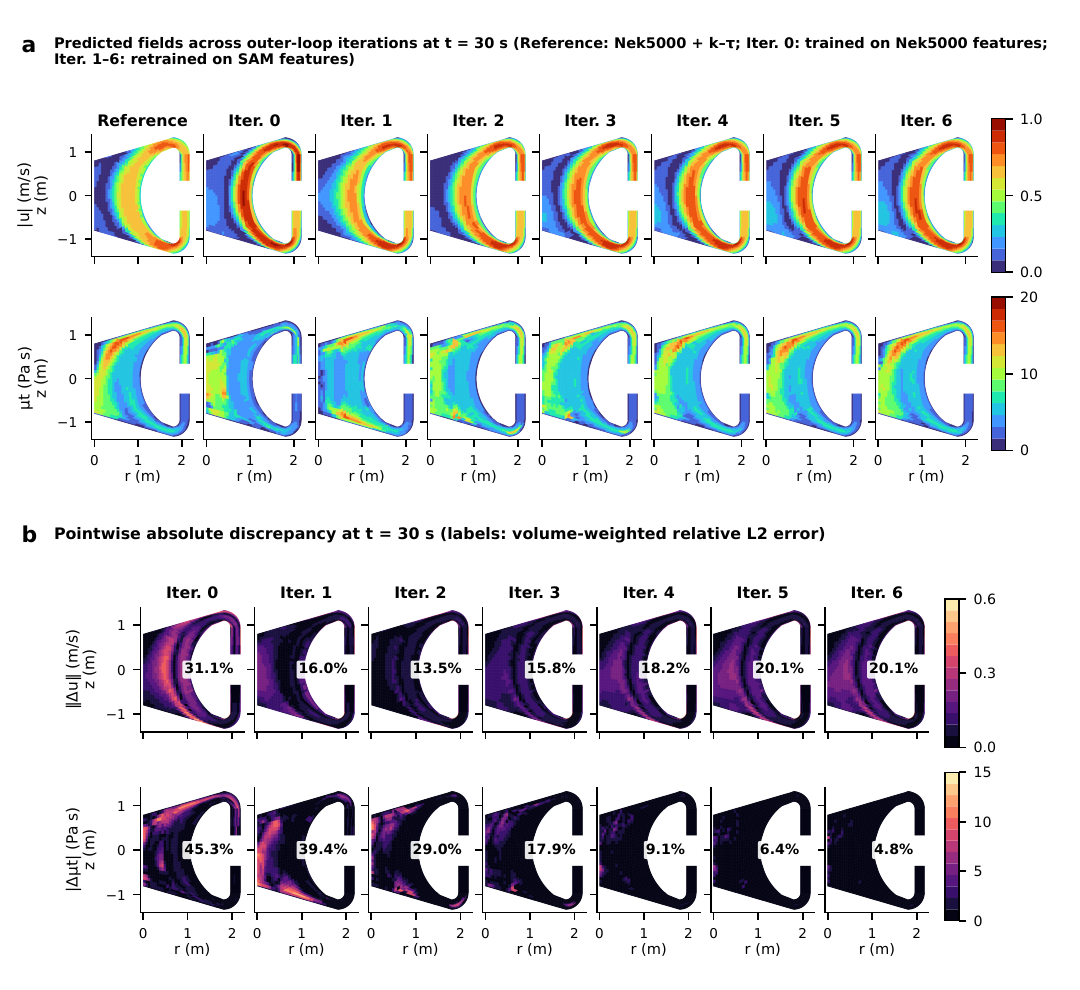}
    \caption{Convergence of the iterative training at $t = 30$~s. \textbf{a}, Predicted velocity magnitude $|\mathbf{u}|$ (top row) and eddy viscosity $\mu_t$ (bottom row) comparing the Nek5000 + $k$--$\tau$ reference against coupled SAM + NN Iterations~0--6 (Iteration~0: network trained on Nek5000 features; Iterations~1--6: retrained on all previous SAM runs). \textbf{b}, Corresponding pointwise absolute discrepancy fields $\|\Delta\mathbf{u}\|_2 = \|\mathbf{u} - \mathbf{u}^*\|_2$ (top row) and $|\Delta\mu_t| = |\widehat{\mu}_t - \mu_t^*|$ (bottom row), where the inset labels report the domain-integrated volume-weighted relative $L_2$ error for each iteration.}
    \label{fig:iteration_progress}
\end{figure}

Figure~\ref{fig:iteration_progress} shows the effect of the iterative training on the coupled SAM fields and pointwise discrepancy maps at $t = 30$~s. In Iteration~0, the network has been trained only on Nek5000 features and suffers strong covariate shift when coupled with SAM. Its eddy-viscosity error is $45.3\%$, and its velocity error is $31.1\%$. Iteration~1, retrained on the SAM features of Iteration~0, reduces the velocity error to $16.0\%$. The following iterations, each retrained on all previous SAM runs, reduce the eddy-viscosity error to $29.0\%$, $17.9\%$, $9.1\%$, $6.4\%$, and $4.8\%$ in Iterations~2--6, close to a tenfold reduction relative to Iteration~0. The velocity error settles at $18$--$20\%$ in Iterations~4--6, close to the $19.6\%$ obtained at $t = 30$~s with prescribed Nek5000 $\mu_t$. The velocity error is not monotonic in the eddy-viscosity error. Its minimum, $13.5\%$, occurs at Iteration~2, because the lower velocity errors of the early iterations reflect a partial cancellation between closure and coarse-mesh discretization errors that disappears as $\widehat{\mu}_t$ approaches the reference.

\begin{figure}[tbp]
    \centering
    \includegraphics[width=0.98\linewidth]{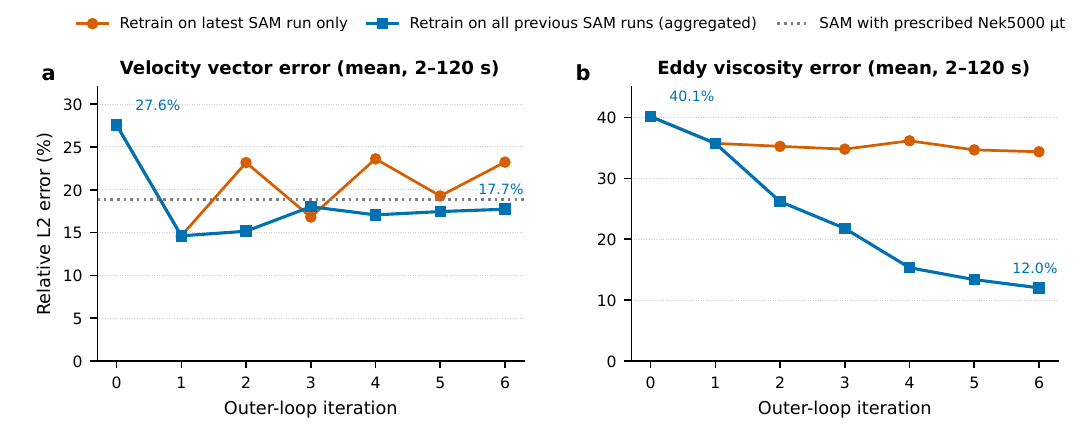}
    \caption{Convergence of the iterative training for the transient. Time-averaged ($2$--$120$~s) volume-weighted relative $L_2$ errors of \textbf{a}, the velocity vector and \textbf{b}, the eddy viscosity versus iteration, for retraining on the latest run (orange) and on all previous runs (blue). The dotted line in \textbf{a} is SAM with prescribed Nek5000 $\mu_t$.}
    \label{fig:iteration_convergence}
\end{figure}

Figure~\ref{fig:iteration_convergence} summarizes the convergence of the iterative training using the time-averaged errors over the full transient. With retraining on the latest run, the coupled velocity error oscillates between about $15\%$ and $24\%$ and the eddy-viscosity error stays near $35\%$. Each network is trained on the states of the previous run, but once coupled it moves the flow to states outside its training data. The flow therefore keeps shifting between iterations, and the eddy-viscosity error stalls. Retraining on all previous SAM runs removes this oscillation. Because the training set keeps every state visited so far, each new network stays accurate on the states produced by earlier networks. The time-averaged eddy-viscosity error then decreases monotonically from $40.2\%$ at Iteration~0 to $12.0\%$ at Iteration~6. The velocity error settles at $17$--$18\%$, close to the $18.8\%$ of SAM with prescribed Nek5000 $\mu_t$.

\begin{table}[tbp]
\centering
\caption{Velocity error of SAM with prescribed Nek5000 $\mu_t$, using first-, second-, and third-order velocity elements on the coarse mesh and first-order elements on the fine mesh. $E_{\mathbf{u}}$ is the volume-weighted relative $L_2$ error of Eq.~\eqref{eq:transient_error}.}
\label{tab:order_mesh}
\footnotesize
\setlength{\tabcolsep}{5pt}
\begin{tabular}{@{} l r r r r @{}}
\toprule
 & \multicolumn{3}{c}{\textbf{Coarse mesh}} & \textbf{Fine mesh} \\
\cmidrule(lr){2-4}\cmidrule(l){5-5}
 & \textbf{First order} & \textbf{Second order} & \textbf{Third order} & \textbf{First order} \\
\midrule
Elements & 1,296 & 1,296 & 1,296 & 4,456 \\
Degrees of freedom & 4,197 & 16,167 & 35,913 & 13,926 \\
\midrule
\multicolumn{5}{@{}l}{\textit{Velocity error $E_{\mathbf{u}}$ at the coarse element centers (\%)}} \\
$t = 10$~s & 20.02 & 20.11 & 19.56 & 7.04 \\
$t = 30$~s & 19.61 & 19.03 & 18.53 & 5.19 \\
$t = 60$~s & 18.85 & 18.21 & 17.77 & 3.80 \\
$t = 120$~s & 17.68 & 17.04 & 16.65 & 2.98 \\
Time average ($2$--$120$~s) & \textbf{18.85} & \textbf{18.38} & \textbf{17.93} & \textbf{4.29} \\
\bottomrule
\end{tabular}
\end{table}

Table~\ref{tab:order_mesh} tests whether element order or mesh resolution sets this common velocity error. SAM is run with prescribed Nek5000 $\mu_t$, so the closure is the same in all cases. The coarse SAM mesh is run with first-, second-, and third-order velocity elements, and the $4{,}456$-element fine mesh with first-order elements. The initial steady state of each run uses the Nek5000 eddy viscosity at $t = 2$~s. For the second and third orders, the SUPG/PSPG element length is divided by the polynomial order, and the third-order case uses hierarchic shape functions with penalty boundary conditions. Because the velocity of these cases varies within each coarse element, all cases are compared at the coarse element centers.

On the coarse mesh, raising the order from one to three increases the degrees of freedom by a factor of $8.6$ and lowers the time-averaged velocity error only from $18.8\%$ to $17.9\%$. On the fine mesh, first-order SAM with $3.3$ times the degrees of freedom of the coarse first-order case reduces the error to $4.3\%$, and to $3.0\%$ at $t = 120$~s. The velocity error of about $18\%$ on the coarse mesh is therefore set by the mesh resolution. Higher-order elements do not remove it, but the fine mesh largely does. The remaining error on the fine mesh, $4.6\%$ when the two velocity fields are compared directly on that mesh, comes from differences between the two solvers. The main difference is SAM's first-order finite elements against Nek5000's seventh-order spectral elements. Because the coarse-mesh velocity error comes from the mesh resolution rather than the closure, a second network could reduce it without refining the mesh. Such a network would correct the coarse-mesh velocity field toward the Nek5000 solution and would be trained with SAM in the loop, as the eddy-viscosity closure is~\cite{um2020solver,jendersie2026hybrid,reza2026recast}.

\FloatBarrier
\subsection{Computational Cost}
\label{sec:cost}

Table~\ref{tab:cost} compares the computational cost of the Nek5000 reference simulations with that of SAM coupled with the final neural closure. The SAM simulations were run on a single core of an AMD Ryzen Threadripper PRO 5995WX workstation. For the steady state, the Nek5000 cost is given for Case~1, the maximum-Reynolds-number case. It ran on 64 cores with a fixed time step of $2.0\times 10^{-5}$. A segment of $1.0\times 10^{6}$ time steps, or 20 time units, took $22{,}700$~s of wall-clock time, which is $22.7$~ms per step. The full duration of 72 time units is extrapolated from this segment. The SAM steady state is the initial condition of the transient, at $Re \approx 1.2\times 10^6$ with $L_0 = 2.0$~m, and is computed by pseudo-transient time stepping. For the 120~s transient, Nek5000 was run on 32 cores with a time step of $1.0\times 10^{-4}$. In terms of CPU core-hours, SAM with the neural closure is about four orders of magnitude cheaper than Nek5000 for the steady state and about three orders of magnitude cheaper for the transient. These timings exclude the one-time cost of the iterative training, which consists of seven coupled SAM runs and the corresponding network retraining.

\begin{table}[h]
\centering
\caption{Computational cost of the Nek5000 reference simulations and of SAM coupled with the mixing-length model and with the final neural closure.}
\label{tab:cost}
\footnotesize
\setlength{\tabcolsep}{4pt}
\begin{tabular}{@{} l l r r r r @{}}
\toprule
\textbf{Simulation} & \textbf{Code} & \textbf{CPU cores} & \textbf{Time steps} & \textbf{Wall-clock time} & \textbf{CPU core-hours} \\
\midrule
Steady state & Nek5000 + $k$--$\tau$ & 64 & $3.6\times 10^{6}$ & $22.7$~h & $1{,}450$ \\
Steady state & SAM + mixing length & 1 & 64 & $2.3$~min & $0.038$ \\
Steady state & SAM + NN & 1 & 63 & $4.5$~min & $0.075$ \\
\midrule
Transient (120~s) & Nek5000 + $k$--$\tau$ & 32 & $1.2\times 10^{6}$ & $10.3$~h & $330$ \\
Transient (120~s) & SAM + mixing length & 1 & 169 & $6.3$~min & $0.11$ \\
Transient (120~s) & SAM + NN & 1 & 169 & $10.5$~min & $0.17$ \\
\bottomrule
\end{tabular}
\end{table}

\FloatBarrier
\section{Conclusions}
\label{sec:conclusions}

In this work, we developed an invariant neural-network eddy-viscosity closure for the coarse-mesh system thermal-hydraulics code SAM and evaluated it on steady-state and transient Molten Salt Fast Reactor (MSFR) flows. The closure inputs are formed on a two-layer neighborhood of edge-adjacent finite elements and pass through a shared-weight embedding network and a quadratic descriptor, so no point-set sampling is needed. The closure satisfies spatial translation, Galilean boost, rotation and reflection, and neighbor-element permutation invariance exactly, and it reconstructs $\widehat{\mu}_t = \rho\widehat{k}\widehat{\tau}$ algebraically at runtime without solving turbulence transport PDEs.

The non-negative eddy viscosity enters SAM's diffusive operator implicitly. This keeps the momentum equations well conditioned and avoids the error amplification that explicit Reynolds-stress source terms cause at high Reynolds numbers. Iterative training against fixed Nek5000 targets removes the discretization-induced covariate shift between seventh-order spectral-element Nek5000 features and coarse $Q_1$ finite-element SAM features. With retraining on all previous runs, the coupled \textit{a posteriori} eddy-viscosity error at $t = 30$~s decreases from $45.3\%$ at Iteration~0 to $4.8\%$ at Iteration~6. Its time average over the transient decreases from $40.2\%$ to $12.0\%$. Retraining on the latest run does not converge.

The coarse SAM mesh has $1{,}296$ $Q_1$ elements and more than two orders of magnitude fewer nodal degrees of freedom than the $4{,}456$-element seventh-order Nek5000 mesh. On this mesh, the neural closure reproduces the single recirculation cell and eddy-viscosity plume of the Nek5000 reference in the held-out steady-state case. In the 120~s pump-coastdown transient, it halves the velocity error of the mixing-length model, from $37.6\%$ to $17.7\%$, and captures the spatial structure and temporal decay of the core eddy-viscosity field. The remaining velocity error is close to the $18.8\%$ of SAM with prescribed Nek5000 $\mu_t$, so it is set mainly by the coarse discretization. The coupled transient takes about $10$ minutes, or $0.2$ core-hours, on a single core, compared with about $330$ core-hours for the Nek5000 reference. The steady state takes about $5$ minutes, compared with an estimated $1{,}450$ core-hours for the maximum-Reynolds-number Nek5000 case, as listed in Table~\ref{tab:cost}.

The present evaluation has several limitations. The reference is a two-dimensional axisymmetric $k$--$\tau$ RANS solution, so the closure is benchmarked against this RANS model rather than validated against experiments or scale-resolving simulations of the three-dimensional MSFR core. Both benchmarks are isothermal, so the closure has not been tested with buoyancy or heat transfer. The transient evaluation covers a single pump-coastdown trajectory from which the training states were also drawn, and the steady-state generalization test uses one held-out Reynolds number. Accuracy degrades outside the training window. In the late extrapolation window, the pooled \textit{a priori} errors are $28.9\%$ for $\mu_t$ and $73.1\%$ for $\tau$, and the coupled eddy-viscosity error reaches $29.7\%$ at $t = 120$~s. The reported speedups also exclude the one-time cost of the iterative training.

Adjoint or differentiable solver-in-the-loop sensitivities~\cite{um2020solver}, or derivative-free ensemble training~\cite{luo2024ensemble}, could accelerate the convergence of the iterative training and enforce smoothness of the learned eddy-viscosity field directly through the discretized momentum residual. Integration with the framework of Almukhametov et al.~\cite{almukhametov2026torchdphys} would allow end-to-end calibration of the closure against thermal-hydraulic responses, so that the training objective accounts for the closure's effect on the coupled flow solution.

\FloatBarrier
\section*{Acknowledgments}
This work is supported by the U.S. Department of Energy Office of Nuclear Energy Distinguished Early Career Program under contract number DE-NE0009468 and the Texas A\&M Institute of Data Science Thematic
Labs Program.

The submitted manuscript has been co-created by UChicago Argonne, LLC, Operator of Argonne National Laboratory (``Argonne''). Argonne, a U.S. Department of Energy Office of Science laboratory, is operated under Contract No. DE-AC02-06CH11357.

\FloatBarrier
\appendix
\section{Neural Closure in the Nonlinear Solver and Its Effect on Convergence}
\label{app:solver}

This appendix describes how the neural closure enters SAM's nonlinear solver and how it affects convergence, in comparison with SAM's mixing-length model and with prescribed Nek5000 $\mu_t$.

\subsection{Coupling of the closure with the solver}
After the SUPG/PSPG finite-element discretization of Section~\ref{sec:sam_zeroeq} and implicit Euler time integration, each time step solves
\begin{equation}
    \mathcal{R}(\mathbf{U}^{n+1}) = \mathbf{M}\,\frac{\mathbf{U}^{n+1}-\mathbf{U}^{n}}{\Delta t} + \mathbf{F}\left(\mathbf{U}^{n+1};\,\mu_t^{n}\right) = \mathbf{0},
    \qquad \mu_t^{n} = \mathcal{F}_\Theta\left(\mathbf{Q}(\mathbf{U}^{n}),\, Re(t_n)\right),
    \label{eq:app_residual}
\end{equation}
where $\mathbf{U}$ holds the nodal velocity and pressure unknowns, $4{,}197$ on the coarse mesh, $\mathbf{M}$ is the mass matrix, and $\mathbf{F}$ contains the convective, pressure, viscous, and stabilization terms. The network is evaluated once per time step, from the accepted state $\mathbf{U}^{n}$, and $\mu_t^{n}$ is held fixed during the Newton iterations
\begin{equation}
    \mathbf{J}\,\delta\mathbf{U}_k = -\mathcal{R}(\mathbf{U}_k), \qquad
    \mathbf{J} = \frac{\mathbf{M}}{\Delta t} + \left.\frac{\partial \mathbf{F}}{\partial \mathbf{U}}\right|_{\mu_t^{n}}.
    \label{eq:app_newton}
\end{equation}
The Jacobian therefore contains no derivative of the network. A fully implicit coupling would instead solve $\mathcal{R}(\mathbf{U}^{n+1};\,\mu_t(\mathbf{U}^{n+1})) = \mathbf{0}$. The two differ by
\begin{equation}
    \mu_t(\mathbf{U}^{n+1}) - \mu_t(\mathbf{U}^{n}) = \frac{\partial \mu_t}{\partial \mathbf{U}}\left(\mathbf{U}^{n+1}-\mathbf{U}^{n}\right) + \mathcal{O}\left(\|\mathbf{U}^{n+1}-\mathbf{U}^{n}\|^2\right) = \mathcal{O}(\Delta t),
    \label{eq:app_lag}
\end{equation}
which is of the same order as the implicit Euler truncation error, so the lagged coupling keeps first-order accuracy in time.

The lagged coupling is used for two reasons. First, the Jacobian $\mathbf{J}$ is assembled by finite differences of the residual with graph coloring, which costs about 43 residual evaluations per Newton iteration on this mesh. An implicit coupling would require evaluating the feature extraction and the network in every residual evaluation, about 110 times per step. Second, $\mu_t$ of an element depends on the velocities of its two-layer neighborhood $\mathcal{N}_0$ of Eq.~\eqref{eq:neighborhood_def}. The derivative $\partial\mu_t/\partial\mathbf{U}$ would therefore couple each momentum equation to velocity unknowns about three element layers away. This would widen the sparsity pattern of $\mathbf{J}$, increase the number of colors, and increase the fill of the LU factorization. With the lagged coupling, $\mathbf{J}$ has the same sparsity and cost as for a constant viscosity, and the finite-difference Jacobian is exact for Eq.~\eqref{eq:app_residual}, so Newton's method converges quadratically.

The final solver settings are: Newton's method with the colored finite-difference Jacobian and an LU factorization; relative and absolute nonlinear tolerances of $10^{-6}$ with at most 15 iterations; adaptive time steps starting at $\Delta t = 0.01$~s with growth factor 1.1 and $\Delta t \le 1$~s for the transient, and growth factor 1.15 and $\Delta t \le 10$~s for the steady state. 


\subsection{Effect on the linear systems}
The eddy viscosity enters the momentum equations through the viscous term. Its Galerkin form is the symmetric bilinear form
\begin{equation}
    a(\mathbf{u}, \mathbf{w}) = \int_\Omega 2\left(\mu + \mu_t\right)\boldsymbol{\varepsilon}(\mathbf{u}) : \boldsymbol{\varepsilon}(\mathbf{w})\,dV, \qquad
    \boldsymbol{\varepsilon}(\mathbf{u}) = \tfrac{1}{2}\left(\nabla\mathbf{u} + \nabla\mathbf{u}^{\mathsf{T}}\right),
    \label{eq:app_visc}
\end{equation}
so its matrix $\mathbf{K}_\mu$ satisfies $\mathbf{v}^{\mathsf{T}}\mathbf{K}_\mu\mathbf{v} = \int_\Omega 2(\mu+\mu_t)\,\boldsymbol{\varepsilon}(\mathbf{v}_h):\boldsymbol{\varepsilon}(\mathbf{v}_h)\,dV \ge 0$ for $\mu_t \ge 0$. Because $\mu_t$ is treated implicitly in this term, the network adds a symmetric positive semidefinite matrix to the momentum block of $\mathbf{J}$. The network output is non-negative by construction and capped at $30$~Pa\,s, so this property holds for any input. The size of the effect is measured by the element P\'eclet number $Pe_h = |\bar{\mathbf{u}}|h/(2\nu_{\mathrm{eff}})$, with $\nu_{\mathrm{eff}} = (\mu+\mu_t)/\rho$, element-averaged velocity $\bar{\mathbf{u}}$, and smallest vertex distance $h$. At $t = 10$~s, $Pe_h$ has a median of $4{,}400$ with molecular viscosity only, $7$ with the network $\mu_t$, and $1.0$ with the mixing-length $\mu_t$, whose median is $45$~Pa\,s and maximum $257$~Pa\,s. The network $\mu_t$ thus turns a strongly convection-dominated momentum operator into a moderately convection-dominated one, which improves the diagonal dominance of $\mathbf{J}$. The network output also enters the SUPG parameter $\tau_m$ through $\nu_{\mathrm{eff}}$; the viscous term dominates $\tau_m$ in $7\%$ of the elements with the network and in $49\%$ with the mixing-length model.

\subsection{Effect on the transient}
Table~\ref{tab:app_closures} compares three closures with the same mesh, boundary conditions, and solver settings. SAM evaluates the mixing-length model inside every residual evaluation, so it is coupled implicitly. The neural closure is lagged by one time step. The prescribed Nek5000 $\mu_t$ is a function of time only and receives no feedback from the flow. In the transient, the three closures need almost the same nonlinear work: $2.2$--$2.7$ Newton iterations per step, no failed solves, and about $16{,}500$--$20{,}000$ residual evaluations. In all three, the time step reaches its maximum of $1$~s at $t = 14$~s, so the network does not limit the time step. At $\Delta t \le 1$~s, the mass term $\mathbf{M}/\Delta t$ in Eq.~\eqref{eq:app_newton} dominates the change of $\mu_t$ from one step to the next. The cost of the network appears outside the solver: feature extraction, network evaluation, and output take $113$~s, or $0.66$~s per step and $18\%$ of the run.

\begin{table}[tbp]
\centering
\caption{Convergence of SAM with three eddy-viscosity closures, with identical mesh, boundary conditions, and solver settings.}
\label{tab:app_closures}
\footnotesize
\setlength{\tabcolsep}{4pt}
\begin{tabular}{@{} l l r r r r @{}}
\toprule
\textbf{Closure} & \textbf{Coupling} & \textbf{Steps} & \textbf{Newton its./step} & \textbf{Failed solves} & \textbf{Residual evals.} \\
\midrule
\multicolumn{6}{@{}l}{\textit{Transient (0--120~s)}} \\
Mixing length & Implicit & 169 & 2.67 & 0 & 19,692 \\
Neural closure (final) & Lagged & 169 & 2.60 & 0 & 19,176 \\
Prescribed Nek5000 $\mu_t$ & None & 169 & 2.24 & 0 & 16,510 \\
\midrule
\multicolumn{6}{@{}l}{\textit{Steady state (pseudo-transient)}} \\
Mixing length & Implicit & 64 & 2.48 & 0 & 6,907 \\
Neural closure (final) & Lagged & 63 & 3.17 & 0 & 8,663 \\
Neural closure (Iteration 0) & Lagged & 198 & 4.70 & 3 & 42,416 \\
\bottomrule
\end{tabular}
\end{table}

\subsection{Effect on the steady state}
The steady states are computed by pseudo-transient continuation with a growing time step, and here the lagged coupling matters. Linearizing one step of Eq.~\eqref{eq:app_residual} about the steady solution $\mathbf{U}^*$ gives the error recursion $\mathbf{e}^{n+1} = \mathbf{A}\,\mathbf{e}^{n}$ with
\begin{equation}
    \mathbf{A} = \left(\frac{\mathbf{M}}{\Delta t} + \mathbf{F}_U\right)^{-1}\left(\frac{\mathbf{M}}{\Delta t} - \mathbf{F}_\mu\,\boldsymbol{\mu}_U\right),
    \qquad
    \mathbf{F}_U = \frac{\partial\mathbf{F}}{\partial\mathbf{U}},\quad
    \mathbf{F}_\mu = \frac{\partial\mathbf{F}}{\partial\mu_t},\quad
    \boldsymbol{\mu}_U = \frac{\partial\mu_t}{\partial\mathbf{U}}.
    \label{eq:app_fixedpoint}
\end{equation}
As $\Delta t \to \infty$, $\mathbf{A} \to -\mathbf{F}_U^{-1}\mathbf{F}_\mu\boldsymbol{\mu}_U$. This is a fixed-point (Picard) iteration between the flow and the closure, which converges only if the spectral radius of $\mathbf{F}_U^{-1}\mathbf{F}_\mu\boldsymbol{\mu}_U$ is less than one. In words, the flow change caused by a closure update, multiplied by the closure change caused by that flow change, must be smaller than the original perturbation. A finite $\Delta t$ keeps part of the previous state through $\mathbf{M}/\Delta t$ and acts as under-relaxation. A closure that is more sensitive to the velocity needs smaller pseudo-time steps, and if it is too sensitive the iteration diverges.

To measure this sensitivity, we define the closure gain
\begin{equation}
    K_\mu = \frac{\|\delta\mu_t\|_V / \|\mu_t\|_V}{\|\delta\mathbf{u}\|_V / \|\mathbf{u}\|_V},
    \label{eq:app_gain}
\end{equation}
the relative change of $\mu_t$ divided by the relative change of the element velocity, in the volume-weighted norm. $K_\mu$ was computed for smooth random velocity perturbations of $0.5\%$, using eight samples with a correlation length of $0.15$~m, applied to SAM states at $t = 10$ and $60$~s. The final network has $K_\mu = 3.0$ and $1.8$, the Iteration~0 network $4.3$ and $2.4$, and a network trained on features computed on the fine Nek5000 mesh $4.5$ and $4.9$. The steady-state convergence follows this order, as listed in Table~\ref{tab:app_closures}. The final network converges in 63 steps without failures. The Iteration~0 network needs 198 steps, $4.7$ Newton iterations per step, and three failed solves, and its output reaches the $30$~Pa\,s cap. Because $K_\mu$ is a norm ratio and not the spectral radius of Eq.~\eqref{eq:app_fixedpoint}, it is an indicator rather than a convergence criterion. The mixing-length model has $K_\mu \approx 2.0$ but is evaluated inside each residual, so its feedback is resolved by Newton's method within each step and does not enter Eq.~\eqref{eq:app_fixedpoint}.

The lag also costs Newton iterations in the steady state: each step starts from a state that is not consistent with the updated $\mu_t$. The final network needs $3.2$ Newton iterations per step and $25\%$ more residual evaluations than the mixing-length model, which needs $2.5$. Finally, the step-to-step relative change in the steady runs with the network levels off near $10^{-5}$ because of the lagged update, so the steady-state tolerance was set to $3\times10^{-5}$.

In summary, embedding the network implicitly in the viscous operator, but evaluating it once per step from the previous state, keeps the Jacobian sparsity and solver cost of a constant-viscosity problem, improves the diagonal dominance of the momentum block, and does not limit the transient time step. Steady-state convergence depends on how sensitive the closure is to the velocity. The final network is the least sensitive and reaches steady state as fast as the mixing-length model, as discussed in Section~\ref{sec:iterative_theory}. If a more sensitive closure is needed, an under-relaxed update $\mu_t^{n} = (1-\omega)\,\mu_t^{n-1} + \omega\,\mathcal{F}_\Theta(\mathbf{Q}(\mathbf{U}^{n}), Re)$ with $0 < \omega < 1$ would reduce the gain of the fixed-point iteration.

\FloatBarrier
\begingroup
\setstretch{1.1}
\bibliographystyle{elsarticle-num}
\bibliography{refs_02}
\endgroup

\end{document}